\documentclass[11pt]{article}
\usepackage{amsmath}

\usepackage{amssymb,color}
\newcommand{\N}{N\raise.7ex\hbox{\underline{$\circ $}}$\;$}

\renewcommand{\theequation}{\thesection.\theequation}
\numberwithin{equation}{section}

\begin{document}

\textwidth 170mm \textheight 230mm \voffset -20mm \hoffset -15mm

\title{Spin 1 particle in  the magnetic monopole potential: \\
nonrelativistic approximation. Minkowski and Lobachevsky spaces}

\maketitle

\author{
O.V. Veko\footnote{Kalinkovichi Gymnasium,
Belarus,vekoolga@mail.ru},
K.V. Kazmerchuk\footnote{Mosyr State
Pedagogical University, Belarus, kristinash2@mail.ru},
E.M. Ovsiyuk\footnote{Mosyr State Pedagogical University, Belarus, ce.ovsiyuk@mail.ru},
V.V. Kisel\footnote{Belarusian State University of Informatics and
Radioelectronics},
A.M. Ishkhanyan\footnote{ Institute for
Physical Research, Armenian Academy of Sciences,
aishkhanyan@gmail.com},
V.M. Red'kov\footnote{B.I. Stepanov Institute of Physics, NAS of Belarus, redkov@dragon.bas-net.by}}

\begin{abstract}

The spin 1 particle is treated in the presence of the Dirac magnetic monopole
 in the Minkowski and Lobachevsky spaces. Separating the variables in the frame
 of the matrix 10-component Duffin-Kemer-Petiau approach (wave equation) and
  making a nonrelativistic approximation in the corresponding radial equations,
  a system of three coupled second order linear differential equations is
  derived for each type of geometry.

For the Minkowski space, the nonrelativistic equations are disconnected
using a linear transformation, which makes the mixing matrix diagonal.
The resultant three unconnected equations involve three routs of a cubic
algebraic equation as parameters. The approach allows extension to
the case of additional external spherically symmetric fields.
The Coulomb and oscillator potentials are considered and for each
of these cases three series of energy spectra are derived.
 A special attention is given to the states
 with minimum value of the total angular momentum.

In the case of the curved background of the Lobachevsky geometry,
 the mentioned linear transformation does not disconnect the nonrelativistic equations in the presence
 of the monopole. Nevertheless, we derive the solution of the problem in the case of minimum total
 angular  momentum. In this case, we additionally involve a Coulomb or oscillator field. Finally,
 considering the case without the monopole field, we show that for both Coulomb and oscillator
 potentials the problem is reduced to a system of three differential equations involving
 a hypergeometric and two general Heun equations.
 Imposing on the parameters of the latter equations a specific requirement, reasonable
 from the physical standpoint, we derive the corresponding energy spectra.

\end{abstract}




%Report to: XVII International
%Conference Foundations \& Advances in Nonlinear Science  September 29 -- October 3, Minsk, 2014. -

\section{Introduction}

The spin has a significant influence on the behavior of the quantum-mechanical particles in the field of the Dirac monopole. To the
present time, however, mainly the particles of spin 0 and 1/2 have gained sufficient attention [1-54].
It should be noted that the case of spin 1/2 Dirac particle reveals very peculiar solutions, which can be associated with the bound states of the particle in the external magnetic monopole field. The existence of similar states for spin 1 particle has been demonstrated in \cite{Red'kov-1988, Olsen-Osland-Wu}. However, other solutions for spin 1 case have not been constructed.

In the present paper we consider the theory of a spin 1 particle in the presence of the  Abelian  monopole background. Our treatment is based on the known matrix Duffin-Kemmer-Petiau formalism extended to the tetrad formalism \cite{Red'kov-2007, Book-1, Book-2, Red'kov-Ovsiyuk-2012}. This approach gives a possibility to achieve a unification with the treatments of the cases of scalar and spin 1/2 particles by means of employing the conventional technique of the Wigner $D$-functions \cite{VMH} (see also \cite{1985-Dray,  1988-Gal'tsov-Ershov, 1994-Weinberg}). Besides, the approach allows one to extend the results straightforwardly to the curved space-time models, in particular, to the case of the Lobachevsky geometry.

We start from the relativistic wave equation for spin 1 particle and arrive at a radial system of equations for 10 functions, which, however, is too complicated for analytical treatment. It then turns out that a progress is achieved in the non-relativistic approximation. Notably, in this limit it is possible to include also the Coulomb or oscillator potentials.

The main lines of the present treatment are the following. The spin 1 particle is considered in the presence of the Dirac magnetic monopole. The separation of the variables is done using the matrix 10-component Duffin-Kemer-Petiau wave equation. For the Minkowski space, the nonrelativistic approximation applied in the radial equations reduces the problem to a system of three second order coupled differential equations, associated with spin 1 particle in the Pauli description. These equations are further disconnected by means of a linear transformation which makes the mixing matrix diagonal. The resultant three unconnected differential equations contain the routs $A_k$, $k=1,2,3$ of a cubic algebraic equation as parameters. The approach permits extension to the case of additional external spherically symmetrical fields. We consider the Coulomb and oscillator potentials. For both cases, we derive three series of energy spectra, $\epsilon = \epsilon \,(A_ {k},\, j, \,n)$,
giving a special attention to the states with minimum value of the total angular  momentum $j$.

Further, we extend the analysis to the case of the curved background of the Lobachevsky geometry.
 Separating the variables in the Duffin-Kemmer-Petiau matrix 10-component equation,
 the problem is again reduced to a system of ten radial equations, however,
 too complicated for analytic treatment. This time also, a progress is achieved in the Pauli
 approximation. Passing to this non-relativistic description, we again derive three coupled second
 order differential equations. However, in contrary to the case of the Minkowski space,
 this time these equations cannot be disconnected in the presence of the monopole.
 Nevertheless, we were able to solve the problem in the case of minimum total
angular  momentum. In this case, it is also possible to take into account the Coulomb
or oscillator fields. For both potentials the solution is written in terms of the hypergeometric functions.

It should be noted that even in the absence of a monopole and even in the nonrelativistic
approximation, the equations for a spin 1 particle in the Lobachevsky
space present a rather complicated mathematical object. We show that in the presence
of a Coulomb or oscillatory potential these equations reduce to a system of three equations involving
 a hypergeometric and two general Heun equations. The Heun equation is
 a direct generalization of the hypergeometric differential equation to the case of
 four regular singular points \cite{Heun-1889,Ronveaux-1995,Slavyanov-Lay-2000}.
 The theory of this equation at present time is far from being complete, and there
 are no known simple methods for analytical description of the corresponding spectrum
  in the general case.  Still, with the use of a special but reasonable from physical
  standpoint requirement, the energy spectrum of the system can be derived in the considered particular cases.

The obtained spectra for a nonrelativistic spin 1 particle in the Lobachebsky geometry background are similar to those encountered in the theory of Scr\"{o}dinger scalar particle in this background (see the details and relevant references in \cite{Red'kov-Ovsiyuk-2012}). Namely, the result reads

for the Coulomb field:
\begin{eqnarray}
\epsilon = - mc^{2} {\alpha ^{2} \over 2 N^{2}} - {\hbar^{2} \over m R^{2} } {N^{2} \over 2} \qquad
\left (\alpha = {e^{2} \over \hbar c}  = {1 \over 137 }  \right )
\;  ,
\end{eqnarray}

for the oscillator potential :
\begin{eqnarray}
\epsilon = \hbar \left ( N \; \sqrt{ {k\over m} +  {\hbar^{2}
\over 4m^{2}R^{4} }} -  {\hbar \over 2mR^{2}} \left(N^{2}+{1\over 4} \right) \right ),
\end{eqnarray}
where $R$ is the curvature radius of the Lobachevsky space, and $N =n+j+1$.
The situation is different for the Minkowski geometry.

\section{Minkowski space. Separation of the variables in the Duffin-Kemer-Petiau equation }

We treat the spin 1 particle in the monopole potential on the basis of the matrix approach in the frame of the tetrad formalism. For the Minkowski space, in notations of \cite{Book-1}, using the spherical tetrad (see \cite{Book-2}), the Duffin-Kemer-Petiau (DKP) equation reads
\begin{eqnarray}
\left [ i\;\beta ^{0}  \partial _{t} \; + \; i \left ( \beta ^{3}
\;
\partial _{r}\; + {1 \over r} \;(\beta ^{1} \; j^{31}  + \beta
^{2} \; j^{32}) \right ) +  {1 \over r}   \Sigma ^{k}_{\theta
,\phi } \; - \; M  \right ]    \Phi (x)  = 0   \; , \label{1.1}
\end{eqnarray}
where
\begin{eqnarray} \Sigma ^{k}_{\theta ,\phi } \; = \;
 i\; \beta ^{1} \; \partial _{\theta } \; + \;
\beta ^{2} \; {i\; \partial_{\phi} \; + \; (i\; j^{12} - k) \cos
\theta \over \sin \theta} \;     . \label{1.1'}
\end{eqnarray}

\noindent  The parameter $k = eg / \hbar c$ is  quantized according to the Dirac rule  $\mid k \mid  =1/2, \,1,\, 3/2,\, 2,\, ...$ \cite{1931-Dirac, 1948-Dirac}. It is convenient to use the cyclic representation for DKP-matrices, for which the third projection of the spin $iJ^{12}$ has a diagonal structure \cite{Book-2, VMH}:
\begin{eqnarray}
iJ^{12} = \left | \begin{array}{cccc}
         0   &   0    &   0     &     0  \\
         0   &  t_{3} &   0     &     0  \\
         0   &   0    &  t_{3}  &     0  \\
         0   &   0    &   0     &    t_{3}
\end{array} \right | \;  , \qquad
 t_{3} = \left | \begin{array}{ccc}
             +1   &   0   &   0  \\
              0   &   0   &   0  \\
              0   &   0   &  -1
\end{array} \right |      \; .
\end{eqnarray}

\noindent The components of the total angular  momentum are given in the spherical tetrad by the formulas \cite{Book-2}:
\begin{eqnarray}
J^{k}_{1} =  l_{1} + { \cos \phi \over \sin \theta} \;(iJ^{12} -
\kappa)  \; , \quad J^{k}_{2} = \;l_{2} + { \sin \phi \over \sin
\theta} \; (iJ^{12} - \kappa) \;  , \quad J^{k}_{3} = l_{3} \;  .
\label{1.2a}
\end{eqnarray}

\noindent In accordance with the general method  \cite{Book-2}, the wave functions of the spin 1 particle with quantum numbers  $(\epsilon , \;j ,\; m )$ are constructed using  the substitution
\begin{eqnarray}
\Phi _{\epsilon jm}(x)  = e^{-i\epsilon t} \; [\; f_{1}(r)\; D_{k}
,\;  f_{2}(r)\; D_{k-1} ,
 \; f_{3}(r) \; D_{\kappa},
\; f_{4}(r) \; D_{k+1},
\nonumber
\\
 f_{5}(r) \; D_{k-1}, \;  f_{6}(r) \; D_{k} , \;
 f_{7}(r) \; D_{k+1}, \;
f_{8}(r) \;D_{k-1} , \; f_{9}(r) \;D_{k} , \; f_{10}(r)\; D_{k+1}
\; ]  \;, \label{1.2b}
\end{eqnarray}

\noindent where the symbol $D_{\sigma}$ stands for the Wigner functions  $D^{j}_{-m, \sigma}  (\phi, \theta, 0)$ \cite{VMH}. Below, when deriving the equations for the ten radial functions $f_{1},\ldots, f_{10}$, we use the following recurrence relations \cite{VMH}:
\begin{eqnarray}
\partial_{\theta } \; D_{k-1} \; =
a \; D_{\kappa-2} - c \; D_{k} \; , \qquad { -m-(k-1) \cos \theta
\over \sin \theta } \; D_{k-1} = -a \; D_{k-2} - c \;  D_{k} \; ,
\nonumber
\\
\partial _{\theta }\; D_{k} \;  = \;
(c \; D_{\kappa-1} -  d \; D_{k+1})\; , \qquad {- m  - k \cos
\theta \over \sin \theta } \; D_{k} = -c \; D_{\kappa-1} - d \;
D_{k+1}\; ,
\nonumber
\\
\partial _{\theta } \; D_{k+1} \; = \;
(d \; D_{k} - b \;  D_{k+2})\; , \qquad {-m-(k+1)\cos \theta \over
\sin \theta } \;  D_{k+1} \; = -d \; D_{k} - b \; D_{k+2}  \; ,
\label{recurent-1}
\end{eqnarray}

\noindent where
\begin{eqnarray}
 a = {1 \over 2}  \sqrt{(j + k
-1)(j - k + 2)}\;  , \qquad b = {1 \over 2}  \sqrt{(j - k -1)(j +
k + 2)} \;,
\nonumber
\\
c = {1 \over 2}  \sqrt{(j + k)(j - k+ 1)} \; , \qquad d = {1 \over
2}  \sqrt{(j - k)(j + k +1)} \; .
\label{parameter-1}
\end{eqnarray}

Eq. (\ref{1.1}) leads to the following system of radial equations:
\begin{eqnarray}
 -i \left ({d \over dr} + {1 \over r} \right ) \; f_{2} -
i {\sqrt{2} c \over r} \;  f_{3} - M \; f_{8} = 0 \; ,
\nonumber
\\
 i  \left ({d
\over dr} + {1 \over r} \right ) \; f_{4} + {i \sqrt{2} d \over r} \;
f_{3}  - M \; f_{10} = 0 \; ,
\nonumber
\\
i \epsilon \;  f_{5} + i \left ( {d \over dr} + {1 \over r} \right ) \; f_{8} +
i {\sqrt{2} c \over r}\; f_{9} - M \; f_{2} = 0 \; ,
\nonumber
\\
i
\epsilon \; f_{7} - i\left  ( {d \over dr} + {1 \over r}  \right ) \; f_{10} - i
{\sqrt{2} d \over r}\;
 f_{9}  - M \; f_{4}= 0 \; ,
\nonumber
\\
-i \epsilon \; f_{2} +  {\sqrt{2} c \over r} \; f_{1}  - M  \;
f_{5} = 0 \; , \quad
 -i \epsilon \; f_{4} +{\sqrt{2} d \over r}
\; f_{1} - M \; f_{7}  = 0 \; ,
\nonumber
\\
- \left ( {d \over dr} + {2 \over r} \right
 ) \; f_{6} - {\sqrt{2} \over r} \; (
c \; f_{5} + d \; f_{7})  - M \; f_{1} = 0 \; ,
\nonumber
\\
 i \epsilon
\; f_{6} + {\sqrt{2} i \over r} \; (- c \; f_{8} + d \; f_{10})
 - M \; f_{3} = 0 \; ,
\nonumber
\\
i{\sqrt{2} \over r} ( c \; f_{2}  - d \; f_{4})  - M \; f_{9} = 0
\; ,\qquad -i \epsilon \; f_{3}  - {d \over dr} \; f_{1} - M \;
f_{6} = 0 \; . \label{1.4}
\end{eqnarray}

The Pauli criterium \cite{1939-Pauli} (see also in \cite{Book-2}) allows $j$ to adopt the values
\begin{eqnarray}
 k = \pm  1/2\;, \qquad \qquad
 j = \mid k\mid  \; ,
 \mid k\mid  + 1 , \,... \,
\nonumber
\\
k= \pm  1, \, \pm  3/2,\, ...  \qquad \qquad
 j = \mid k\mid - 1,\; \mid k\mid
,\; \mid k\mid  + 1,\; ... \label{1.5}
\end{eqnarray}

\noindent We have separated here the values $k = \pm 1/2$ since these need a different discussion, see below. Note also that the states with  $j= \mid k\mid $ and  $j= \mid k\mid  - 1 $ should be treated with special caution, because the system (\ref{1.4}) is meaningless in these cases.

Consider the states with $j = \mid k\mid -1$. For $j = 0$ (that is if  $k = \pm  1$), the corresponding wave function does not depend on the angular variables $\theta ,\phi $. If $j = 0$ and $k = + 1$, the initial substitution is
\begin{eqnarray}
\Phi^{(0)} (t,r) = \; e^{-i\epsilon t} \; ( \; 0 , \;\; \; f_{2}
,\;\;  0 , \;\;  0 , \;\; f_{5}, \;\; 0 ,\; 0 ; \;\; f_{8} , \;\;
0 , \;\; 0 \; ) \; . \label{1.6a}
\end{eqnarray}

\noindent It is readily checked that the angular operator $\Sigma
_{\theta ,\phi }$ acts  on  $\Phi ^{(0)}$ as a zero operator:
 $\Sigma _{\theta ,\phi }  \Phi ^{(0)}  = 0$. This results in only three
 nontrivial radial equations:
\begin{eqnarray}
i \; \epsilon \; f_{5} \; + \; i \;\left  (\;  {d \over dr}\; + \;{1
\over r}\; \right ) \;f_{8} - M \; f_{2}\; =  0  \; ,
\nonumber
\\
-\; i\; \epsilon  f_{2}\; -\; M \; f_{5} \; = 0  \; , \;\;\;
  -\; i\; \left (\; {d \over dr}\; +\;  {1 \over r} \right )\;
 f_{2} - M \; f_{8} \; = 0     \; ,
\label{1.6b}
\end{eqnarray}

\noindent from which we get
\begin{eqnarray}
f_{5} =  -  i  {\epsilon \over M}  f_{2} \; , \quad
f_{8} =  -  {i \over M}  \left (  {d \over dr}  +  {1 \over r} \right  ) f_{2} \; , \quad
\left ( {d^{2} \over dr^{2}}  +  \epsilon ^{2}    -  M^{2} \right )  F_{2}  = 0 \; , \quad
f_{2}(r)={1 \over r}F_{2}(r) \; .
\label{1.6c}
\end{eqnarray}

\noindent The equation for $F_2$ is exactly the same as the one that appears in the case of the electron in the same situation \cite{Book-2}.

The case $j= 0$,  $k = - 1$ is treated in the similar way. The proper initial substitution this time is
\begin{eqnarray}
\Phi ^{(0)}(t,r)\; =\; e^{-i\epsilon t} \; (\; 0, \;\; 0,\;\;
0,\;\; f_{4}\; ,\;\; 0, \;\; 0, \;\; f_{7}, \;\;   0, \;\; 0, \;\;
f_{10}\; ) ,
\label{1.7a}
\end{eqnarray}

\noindent and the corresponding radial equations read
\begin{eqnarray}
i \; \epsilon\; f_{7}\; - i\;   \left ({d \over dr} \; + \; {1 \over r} \right )
\; f_{10} \;-\; M \;f_{4} = 0 \;  ,
\nonumber
\\
-\; i \; f_{4} \; - \;M \; f_{7} = 0 \; , \qquad
 i\;  \left ({d \over dr} \;+\; {1 \over r} \right )\; f_{4} \; - M \; f_{10} = 0    \; .
\label{1.7b}
\end{eqnarray}

\noindent  As  a result, we derive
\begin{eqnarray}
f_{7} = -i{ \epsilon \over M}  f_{4}  \; , \quad
f_{10}  = {i \over M }  \left  ({d \over dr} + {1 \over r} \right )  f_{2} \; , \quad
\left ( {d^{2} \over dr^{2}}  +  \epsilon ^{2}  - M^{2}\right  )
F_{4} = 0 \; , \quad  f_{4} = {1 \over r}  F_{4}\;  \; .
\label{1.7c}
\end{eqnarray}

Now, consider the case of minimum values of  $j = \mid k\mid -1$ with  $ k= \pm 3/2 , \; \pm  2 , \; \ldots $. First, we assume that $k$ is positive. In this case we should start with the substitution
\begin{eqnarray}
k \ge  3/2  , \qquad \Phi ^{(0)} = e^{-i\epsilon t}\;  ( \;0 ,\;
f_{2}\; D_{k-1} ,\; 0 ,\; 0 ; f_{5} \; D_{k-1} ,\; 0 , \; 0 ;\;
f_{8}\; D_{k-1} ,\; 0 ,\; 0\; ) \; .
\label{1.8}
\end{eqnarray}

\noindent Using the recurrence relations \cite{VMH}
\begin{eqnarray}
\partial _{\theta}\; D_{\kappa-1} =  \sqrt{{k-1 \over 2}}\;
 D_{k-2} \; , \qquad
{-m - (k-1) \cos \theta \over \sin \theta} \; D_{k-1} =
 -  \sqrt{{k-1 \over 2}}\; D_{k-2}   \; ,
 \label{1.888}
\end{eqnarray}
\noindent we derive $ \Sigma _{\theta ,\phi } \Phi ^{(0)} = 0\; $.
Therefore, the system of the radial equations for the functions  $f_{2},\; f_{5},\; f_{8}$
coincides with Eqs. (\ref{1.6b}).
The case  $j=\mid k\mid -1$  with negative $k$ is considered in the similar way:
\begin{eqnarray}
k \leq -3/2, \qquad  \Phi ^{(0)} = e^{-i\epsilon t}\;
 (\; 0 ,\; 0 ,\; 0 ,\;\; f_{4}\; D_{k+1} ,\; \; 0 ,\; 0 ,
\;\; f_{7}\; D_{k+1} ,\; 0 ,\; 0 ,\;\; f_{10}\; D_{k+1}\;\; ) \; .
\label{1.9}
\end{eqnarray}

\noindent Here again holds the identity $\Sigma _{\theta ,\phi } \Phi ^{(0)} = 0$, so the radial system coincides with Eqs. (\ref{1.7b}).

Thus, for all states with minimum values of $j$, $j = \mid k\mid - 1$, we derive simple sets of radial equations, which have straightforward analytical solutions of exponential form. However, this is not the case for the states with greater $j$. In the latter case we have a complicated system of ten radial equations, the general solution of which is not known. However, as shown in the next section, the system is exactly solved in the nonrelativistic approximation.

\section{The Pauli approximation}

We employ the method of deriving non-relativistic equations used in \cite{Book-2}. First, we eliminate from Eqs. (\ref{1.4}) the non-dynamical variables $f_1$ and $f_{8,9,10}$. The resultant six equations in the more symmetric notations $f_{2,3,4} \to \Phi_{1,2,3}$ and $f_{5,6,7} \to E_{1,2,3}$ read
\begin{eqnarray}
i \epsilon M  E_{1}  + i  \left ( {d \over dr} + {1 \over r} \right ) \left [
-i \left ({d \over dr} + {1 \over r} \right )  \Phi_{1} - i {\sqrt{2} c \over r}
\Phi_{2} \right ] + i {\sqrt{2} c \over r} \left [i{\sqrt{2} \over
r} ( c  \Phi_{1}  - d  \Phi_{3}) \right  ] - M^{2}  \Phi_{1}  = 0
\; ,
\nonumber
\\
\hspace{-2mm} i \epsilon M E_{2} + {\sqrt{2}i \over r} \left  [- c
 \left ( -i \left({d \over dr} + {1 \over r} \right)  \Phi_{1} - i {\sqrt{2} c \over r}   \Phi_{2}  \right ) + d \;
\left ( i \left({d \over dr} + {1 \over r} \right)  \Phi_{3} + {i \sqrt{2} d
\over r}  \Phi_{2}  \right ) \right ]
 - M^{2}  \Phi_{2} = 0 \; ,
\nonumber
\\
i \epsilon M E_{3} - i \left ( {d \over dr} + {1 \over r} \right ) \left [ i
\left ({d \over dr} + {1 \over r} \right )  \Phi_{3} + {i \sqrt{2} d \over r}
\Phi_{2} \right  ] - i {\sqrt{2} d \over r}
 \left [i{\sqrt{2} \over r} ( c  \Phi_{1}  - d  \Phi_{3}) \right ]  - M^{2} \Phi_{3}= 0 \; ,
\nonumber
\end{eqnarray}
\begin{eqnarray}
-i \epsilon M  \Phi_{1} +  {\sqrt{2} c \over r}  \left [ - \left ( {d
\over dr} + {2 \over r} \right )  E_{2} - {\sqrt{2} \over r}  ( c \;
E_{1} + d  E_{3}) \right  ]  - M^{2}  E_{1} = 0 \; ,
\nonumber
\\
-i \epsilon M \Phi_{2}  - {d \over dr}  \left [ - \left ( {d \over dr} +
{2 \over r}\right  )  E_{2} - {\sqrt{2} \over r}  ( c  E_{1} + d \; E_{3}
) \right ] - M ^{2} E_{2} = 0 \; ,
\nonumber
\\
-i \epsilon M  \Phi_{3} +{\sqrt{2} d \over r}   \left [ - \left ( {d
\over dr} + {2 \over r} \right )  E_{2} - {\sqrt{2} \over r}  ( c  E_{1}
+ d
 E_{3} ) \right ] - M^{2}  E_{3} = 0  \; .
\label{2.5}
\end{eqnarray}

\noindent Separating the rest energy with the help of the formal change $\epsilon = M + E$ and further regrouping the equations in appropriate pairs, we get the following three systems
\begin{eqnarray}
i M^{2} E_{1}    + iE M E_{1}   +
\left  ({d ^{2}\over dr^{2} } + {2 \over r} {d \over dr}\right  )
 \Phi_{1} + {\sqrt{2} c \over r}   {d \over dr}  \Phi_{2}
- {2 c \over r^{2} } ( c  \Phi_{1}  - d \; \Phi_{3})  - M^{2} \;
\Phi_{1}  = 0  ,
\nonumber
\\
-i M^{2} \Phi_{1}  - i E  M  \Phi_{1}  - {\sqrt{2} c \over r}  \left (
{d \over dr} + {2 \over r} \right )  E_{2} - {2 c \over r^{2}}  ( c
E_{1} + d  E_{3})   - M^{2}  E_{1} = 0 \;  ,
\label{2.111}
\end{eqnarray}
\begin{eqnarray}
i M^{2} E_{2}   +  i E  M E_{2} - {\sqrt{2}  c  \over r}
 \left ( {d \over dr} + {1 \over r} \right )  \Phi_{1} -  {2 c^{2} \over r^{2} }   \Phi_{2}
  - { \sqrt{2} d \over r } \left  ({d \over dr} + {1 \over r} \right
  ) \; \Phi_{3} - {2  d^{2} \over r^{2} }  \Phi_{2}
 - M^{2}  \Phi_{2} = 0  ,
\nonumber
\\
-i M^{2} \Phi_{2}  - i  E   M \Phi_{2}   + {d \over dr} \left ( {d \over
dr} + {2 \over r} \right  ) \; E_{2} + {d \over dr}  {\sqrt{2} \over r}  (
c  E_{1} + d E_{3} )  - M ^{2} E_{2} = 0   ,\qquad
\label{2.222}
\end{eqnarray}
\begin{eqnarray}
i M^{2} E_{3}   + iE  M E_{3} + \left ({d^{2} \over dr^{2}} + {2 \over
r}{d \over dr} \right )  \Phi_{3} + { \sqrt{2} d \over r} {d \over dr}
\Phi_{2} + { 2 d \over r^{2} } ( c  \Phi_{1}  - d \; \Phi_{3})   -
M^{2}  \Phi_{3}= 0  \; ,
\nonumber
\\
-i M^{2} \Phi_{3}  - i E  M  \Phi_{3}    - {\sqrt{2} d \over r} \left (
{d \over dr} + {2 \over r}\right  )  E_{2} - { 2d \over r^{2}}  ( c
E_{1} + d
 E_{3} ) - M^{2}  E_{3} = 0  \; .
\label{2.333}
\end{eqnarray}

%
%$$
% M^{2} (\Psi_{1} - \psi_{1})     + E  M (\Psi_{1} - \psi_{1})    +
% \left ({d ^{2}\over dr^{2} } + {2 \over r} {d \over dr}  \right )   (\Psi_{1} + \psi_{1})  +
% {\sqrt{2} c \over r}   {d \over dr}  (\Psi_{2} + \psi_{2})
%-
%$$
%$$
%- {2 c \over r^{2} } [ c  (\Psi_{1} + \psi_{1})   - d \; (\Psi_{3}
%+ \psi_{3})  ]   - M^{2} \; (\Psi_{1} + \psi_{1})   = 0  \; ,
%$$
%$$
%-i M^{2} (\Psi_{1} + \psi_{1})   -  iE  M  (\Psi_{1} + \psi_{1})
%+ i {\sqrt{2} c \over r}  \left ( {d \over dr} + {2 \over r}\right  )
%(\Psi_{2} - \psi_{2})   +
%$$
%$$
%+ i {2 c \over r^{2}}  [ c  (\Psi_{1} - \psi_{1}) + d  (\Psi_{3} -
%\psi_{3})   ]   + i M^{2} (\Psi_{1} - \psi_{1})   = 0  \; ,
%$$
%
%\noindent and regrouping the terms we arrive at the equations
%\begin{eqnarray}
%-  M^{2}  \psi_{1}     + E  M (\Psi_{1} - \psi_{1})    +
%\left  ({d ^{2}\over dr^{2} } + {2 \over r} {d \over dr}  \right )   (\Psi_{1} + \psi_{1})  +
% {\sqrt{2} c \over r}   {d \over dr}  (\Psi_{2} + \psi_{2})
%\nonumber
%\\
%- {2 c \over r^{2} } [ c  (\Psi_{1} + \psi_{1})   - d \; (\Psi_{3}
%+ \psi_{3})  ]   = 0  \; ,
%\nonumber
%\\
%- M^{2}  \psi_{1}   -  E  M  (\Psi_{1} + \psi_{1})  +  {\sqrt{2} c
%\over r}  \left ( {d \over dr} + {2 \over r} \right  )  (\Psi_{2} - \psi_{2})
%\nonumber
%\\
%+  {2 c \over r^{2}}  [ c  (\Psi_{1} - \psi_{1}) + d  (\Psi_{3} -
%\psi_{3})   ]     = 0  \; . \label{2.8c}
%\end{eqnarray}

Introducing the big $\Psi_{j}$ and small $\psi_{j}$ constituents via the linear relations
\begin{eqnarray}
\Psi_{j} =  \Phi_{j} + i E_{j} , \qquad \psi_{j} =  \Phi_{j} - i E_{j} \; ,
\label{2.6}
\end{eqnarray}
one recovers from these systems the three equations of the Pauli approximation {\cite{Book-2}. Indeed, e.g., rewriting the first pair (\ref{2.111}) in terms of the big and small components and subtracting the second equation from the first one, we obtain the following equation involving only the big components:
\begin{eqnarray}
\left (  {d ^{2}\over dr^{2} } + {2 \over r} {d \over dr}  + 2E  M
\right )    \Psi_{1} -{2\sqrt{2} c \over r^{2} } \Psi_{2}
-{4c^{2} \over r^{2} } \Psi_{1}  = 0 \; ,
\label{2.100}
\end{eqnarray}

\noindent which represents one of the equations in the nonrelativistic approximation. The second and the third equations of this approximation,  derived in the same manner from systems (\ref{2.222}) and (\ref{2.333}), respectively, read
\begin{eqnarray}
\left ( {d^{2} \over dr^{2}}  + {2 \over r} {d \over dr} +  2E  M
 \right )  \Psi_{2} -
 { 2(c^{2} +d^{2} +1) \over r^{2} }    \Psi_{2}
- { 2\sqrt{2} c \over r^{2} }  \Psi_{1} -  { 2 \sqrt{2} d \over
r^{2} }  \Psi_{3} = 0 \; .
\label{2.200}
\end{eqnarray}

\begin{eqnarray}
\left (  {d^{2} \over dr^{2}} + {2 \over r}{d \over dr} + 2E M
\right )  \Psi_{3} -{2 \sqrt{2} d \over r^{2} } \Psi_{2} - {4d^{2}
\over r^{2} } \Psi_{3} = 0\; .
\label{2.300}
\end{eqnarray}

To make a non-relativistic approximation in the case of minimum $j$, one eliminates the nondynamical component $f_{8}$ from the system (\ref{1.6b}) and changes to the big and small components $\Psi_{j}$ and $\psi_{j}$ to obtain, instead of (\ref{2.100}), a simpler equation:
\begin{eqnarray}
\left (  {d ^{2} \over dr^{2} } + {2 \over r} {d \over dr}  + 2EM
\right ) \Psi_{1}= 0\; ,
\label{5.4}
\end{eqnarray}

\noindent from which we derive the nonrelativistic solution
\begin{eqnarray}
\Psi_{1} ={1 \over r} f_{2}(r), \quad f_{2} = e^{\pm  \sqrt{-2ME} r} .
\label{5.5}
\end{eqnarray}

\noindent Thus, the specific relativistic solution that could be associated with a bound state at $0< \epsilon < M$ (we recall that  $\epsilon = M + E$), in the nonrelativistic approximation reads
\begin{eqnarray}
\Psi_{1} = {e^{-\sqrt{-2EM} \; r}  \over r} \; ,
\label{5.6}
\end{eqnarray}
\noindent which is associated with a bound state as well.

\section{Solutions of the radial Pauli equations in the general case}

%No we are to examine radial system in the Pauli approximation
%
% \begin{eqnarray}
%\left (  {d ^{2}\over dr^{2} } + {2 \over r} {d \over dr}  + 2E  M
%\right )    \Psi_{1} -{2\sqrt{2} c \over r^{2} } \Psi_{2}
%-{4c^{2} \over r^{2} } \Psi_{1}  = 0 \; ,
%\nonumber
%\\
%\left ( {d^{2} \over dr^{2}}  + {2 \over r} {d \over dr} +2E  M
% \right )  \Psi_{2} -
% { 2(c^{2} +d^{2} +1) \over r^{2} }    \Psi_{2}
%- { 2\sqrt{2} c \over r^{2} }  \Psi_{1} -  { 2 \sqrt{2} d \over
%r^{2} }  \Psi_{3} = 0 \; ,
%\nonumber
%\\
%\left (  {d^{2} \over dr^{2}} + {2 \over r}{d \over dr} + 2E M
%\right )  \Psi_{3} -{2 \sqrt{2} d \over r^{2} } \Psi_{2} - {4d^{2}
%\over r^{2} } \Psi_{3} = 0 \; . \label{3.1}
%\end{eqnarray}
%

To examine the radial system of the Pauli approximation in the general case of non-minimum angular
 momentum, it is convenient to introduce  the notation
\begin{eqnarray}
\bar{\Delta}={1 \over 2} r^{2} \left (  {d ^{2}\over dr^{2} } + {2 \over r} {d
\over dr}  + 2E  M \right ),
\end{eqnarray}

\noindent so that the system (\ref{2.100})-(\ref{2.300}) is written in the matrix form as
\begin{eqnarray}
\bar{\Delta} \; \Psi = \bar{A} \Psi \; ,\qquad \bar{A} = \left |
\begin{array}{ccc}
2c^{2 }  & \sqrt{2}c  &  0 \\
\sqrt{2} c &  (c^{2}+d^{2} +1)       &    \sqrt{2} d \\
0 & \sqrt{2} d & 2d^{2}
\end{array} \right |, \qquad \Psi = \left | \begin{array}{c}
\Psi_{1} \\ \Psi_{2} \\ \Psi_{3}
\end{array} \right | .
\label{3.2}
\end{eqnarray}

\noindent We recall that the nonrelativistic wave function is defined by the relations
\begin{eqnarray}
 \Phi _{\epsilon jm} (x)  = e^{-i\epsilon t} \;
 \left | \begin{array}{l}
 \Psi_{1} (r) \; D_{k-1} \\
  \Psi_{2} (r)  \; D_{\kappa} \\
\Psi_{3} (r) \; D_{k+1}
\end{array} \right | = e^{-i\epsilon t} \;
 \left | \begin{array}{l}
 (\Phi_{1} + iE_{1} )\; D_{k-1} \\
  (\Phi_{2} + i E_{2} )  \; D_{\kappa} \\
 (\Phi_{3}  + i E_{3} )\; D_{k+1}
\end{array} \right | .
\label{3.2c}
\end{eqnarray}

To solve the system (\ref{3.2}), we diagonalize the matrix $\bar{A}$. The equation for determining the corresponding transformation matrix $S$ and the diagonal elements $A_{1}, A_{2}, A_{3}$ reads
\begin{eqnarray}
\left | \begin{array}{ccc}
2c^{2 }  & \sqrt{2}c  &  0 \\
\sqrt{2} c &  (c^{2}+d^{2} +1)       &    \sqrt{2} d \\
0 & \sqrt{2} d & 2d^{2}
\end{array} \right |
\left | \begin{array}{ccc}
s_{11} & s_{12} & s_{13} \\
s_{21} & s_{22} & s_{23} \\
s_{31} & s_{32} & s_{33}
\end{array} \right |
  = \left | \begin{array}{ccc}
s_{11} & s_{12} & s_{13} \\
s_{21} & s_{22} & s_{23} \\
s_{31} & s_{32} & s_{33}
\end{array} \right |
 \left | \begin{array}{ccc}
A_{1} & 0 & 0 \\
0 & A_{2}& 0 \\
0 & 0 & A_{3}
\end{array} \right |
\label{3.3a}
\end{eqnarray}

%
%\noindent The corresponding equations can be divided into three groups:
%\begin{eqnarray}
%2c^{2} s_{11} + \sqrt{2} c s_{21}  = s_{11} A_{1}\;,
%\nonumber
%\\
%\sqrt{2} c  s_{11} +(c^{2}+d^{2} +1) s_{21} + \sqrt{2}d s_{31} =
%s_{21} A_{1}\; ,
%\nonumber
%\\
%\sqrt{2} d s_{21} + 2d^{2} s_{31} = s_{31} A_{1}\; ;
%\nonumber
%\\[3mm]
%2c^{2} s_{12} + \sqrt{2} c s_{22}  = s_{12} A_{2}\; ,
%\nonumber
%\\
%\sqrt{2} c  s_{12} +(c^{2}+d^{2} +1) s_{22} + \sqrt{2}d s_{32} =
%s_{22} A_{2}\; ,
%\nonumber
%\\
%\sqrt{2} d s_{22} + 2d^{2} s_{32} = s_{32} A_{2} \; ;
%\nonumber
%\\[3mm]
%2c^{2} s_{13} + \sqrt{2} c s_{23}  = s_{13} A_{3}\; ,
%\nonumber
%\\
%\sqrt{2} c  s_{13} +(c^{2}+d^{2} +1) s_{23} + \sqrt{2}d s_{33} =
%s_{23} A_{3}\; ,
%\nonumber
%\\
%\sqrt{2} d s_{23} + 2d^{2} s_{33} = s_{33} A_{3}\; . \label{3.3b}
%\end{eqnarray}

\noindent This equation can be divided into three groups of the same structure, each of which presents a set of three homogeneous linear equations:
\begin{eqnarray}
\left | \begin{array}{ccc}
(2c^{2}-A_{1} ) &  \sqrt{2} c & 0 \\
\sqrt{2} c  & (c^{2}+d^{2} +1 - A_{1} ) & \sqrt{2}d \\
0 & \sqrt{2} d &  (2d^{2} -A_{1})
\end{array} \right |
\left |  \begin{array}{c} s_{11} \\ s_{21} \\ s_{31}
\end{array} \right | = 0 \;  ,
\label{3.3a}
\\
\left | \begin{array}{ccc}
(2c^{2}-A_{2} ) &  \sqrt{2} c & 0 \\
\sqrt{2} c  & (c^{2}+d^{2} +1 - A_{2} ) & \sqrt{2}d \\
0 & \sqrt{2} d &  (2d^{2} -A_{2})
\end{array} \right |
\left |  \begin{array}{c} s_{12} \\ s_{22} \\ s_{32}
\end{array} \right | = 0 \; ,
\label{3.3b}
\\
\left | \begin{array}{ccc}
(2c^{2}-A_{3} ) &  \sqrt{2} c & 0 \\

\sqrt{2} c  & (c^{2}+d^{2} +1 - A_{3} ) & \sqrt{2}d \\
0 & \sqrt{2} d &  (2d^{2} -A_{3})
\end{array} \right |
\left |  \begin{array}{c} s_{13} \\ s_{23} \\ s_{33}
\end{array} \right | = 0 \; .
\label{3.3c}
\end{eqnarray}

\noindent To have non-trivial solutions, the determinants of the involved matrixes should vanish, hence, we conclude that the diagonal elements $A_{1,2,3}$ are the roots of the following cubic algebraic equation:
\begin{eqnarray}
\det \left | \begin{array}{ccc}
(2c^{2}-A ) &  \sqrt{2} c & 0 \\
\sqrt{2} c  & (c^{2}+d^{2} +1 - A ) & \sqrt{2}d \\
0 & \sqrt{2} d &  (2d^{2} -A)
\end{array} \right | = 0 \; ,
\end{eqnarray}
or, equivalently,
\begin{eqnarray}
 A^{3} + r A^{2} + s A + t = (A -A_1)(A -A_2)(A -A_3)=0 .
\label{3.5a}
\end{eqnarray}

Eliminating the $A^2$ term in this equation by means of the substitution $A=B - r/3$, we get a reduced cubic equation:
\begin{eqnarray}
B^{3} + p B + q = 0\; ,\quad
p= {3s - r^{2} \over 3}\;, \quad q = {2r^{3} \over 27 } - {rs
\over  3} + t \; .
\label{3.6c}
\end{eqnarray}

\noindent As it is known, the properties of the roots of a cubic equation
depend on the sign of its discriminant \cite{Abramowitz-Stegun}. For a reduced cubic equation the discriminant is defined as
\begin{eqnarray}
D = \left({p\over 3} \right)^{3} + \left({q\over 2} \right)^{2},
\label{3.8a}
\end{eqnarray}
\noindent and the negative sign of $D$, $D<0$, warrants that all three roots are real. After some simplifications, we get that
\begin{eqnarray}
p= - \left ( j(j+1) -{3 \over 4} k^{2}  + {1 \over 3} \right ) < 0
\; ,\qquad q =  - \left (   {1 \over 3} j(j+1) +{2 \over 27}
\right ) < 0 \; .
\label{3.7b}
\end{eqnarray}

\noindent Expanding now the expression (\ref{3.8a}) for the discriminant with this $p$ and $q$, it is readily checked that always $D<0$. Hence, the reduced cubic equation has only real roots. It is convenient to write the roots of this equation in the standard trigonometric form \cite{Abramowitz-Stegun}:
\begin{eqnarray}
B_{i} =
 2 \sqrt{-{p\over 3}} \cos \left( {1\over 3} \arccos \left({3\;q \over 2\;p}\sqrt{{-{3\over p}}} \right)
 + (i-1){2 \pi \over 3} \right), \quad i=1,2,3\; .
\end{eqnarray}

Thus, the roots $A_{1,2,3}$ are real. Since the free term $t$ in (\ref{3.5a}) is negative, hence, $A_1A_2A_3>0$, for the signs of these roots only the following variants are possible:
\begin{eqnarray}
\textrm{sign}(A_1,A_2,A_3) =(+,+,+) \quad \textrm{or} \quad (-,-,+).
\label{A123singns}
\end{eqnarray}
The numerical calculations for different $j$
additionally reveal that the roots $A_{1,2,3}$ all are positive.

%With $A=B-r/3$ and $r=-(3M +1)<0$ (since $M>0$, see Eq. (\ref{MN})),
%we see that the roots $A_{1,2,3}$ are always positive.

%
%The analytical form of these roots are written in a standard way
%introducing two auxiliary quantities:
%$$
%\rho =\sqrt{-{p^{3} \over 27}}\; , \qquad \cos \phi = -{q \over 2
%\rho }\; ;
%%\eqno(3.10a)
%$$
%
%\noindent then  the roots of reduced equation are given by
%\begin{eqnarray}
%B_{1} = 2 \rho^{1/3} \;  \cos {\phi \over 3} = 2 \sqrt{-{p\over
%3}} \;\; \cos {\phi \over 3}  \; ,
%\nonumber
%\\
%B_{2} = 2 \rho^{1/3} \;  \cos [ {\phi \over 3} + {2 \pi \over 3}]
%=
% 2 \sqrt{-{p\over 3}}   \;\; \cos ( {\phi \over 3}+  {2 \pi \over 3})   \; ,
%\nonumber
%\\
%B_{3} = 2 \rho^{1/3} \;  \cos [ {\phi \over 3} + {4 \pi \over 3}]=
% 2 \sqrt{-{p\over 3}}  \;\;  \cos ( {\phi \over 3}-  {2 \pi \over 3})\; ,
%\label{3.10b}
%\end{eqnarray}
%
%\noindent where
%\begin{eqnarray}
%\sqrt{-{p \over 3} } =  \sqrt{{j(j+1)-k^{2} \over 3} + {k^{2}
%\over 12 } + {1 \over 9}}\; ,
%\nonumber
%\\
%\rho = \sqrt{-{p^{3} \over 27}} =
% \left ( {j(j+1) -k^{2} \over 3}  +{k^{2}  \over 12} k^{2}  + {1 \over 9} \right )^{3/2} ,
%\nonumber
%\\
%\cos \phi = -{q /2\over  \rho} = { \left (   {1 \over 6} j(j+1)
%+{1 \over 27}  \right )\over \left ( {j(j+1)-k^{2} \over 3} +
%{k^{2} \over 12 } + {1 \over 9} \right )^{3/2}} \; . \label{3.10c}
%\end{eqnarray}
%

\section{Explicit form of the three types of solutions}

%
%Let us turn back to
% (\ref{3.3b}). The matrix  $S_{ij}$ is determined by 9 equations
% (in fact, each sub-system contains only two independents equations) :
%\begin{eqnarray}
%2c^{2} s_{11} + \sqrt{2} c s_{21}  = s_{11} A_{1}\; ,
%\nonumber
%\\
%\sqrt{2} c  s_{11} +(c^{2}+d^{2} +1) s_{21} + \sqrt{2}d s_{31} =
%s_{21} A_{1}\; ,
%\nonumber
%\\
%\sqrt{2} d s_{21} + 2d^{2} s_{31} = s_{31} A_{1}\; ;
%\nonumber
%\\[3mm]
%2c^{2} s_{12} + \sqrt{2} c s_{22}  = s_{12} A_{2}\;,
%\nonumber
%\\
%\sqrt{2} c  s_{12} +(c^{2}+d^{2} +1) s_{22} + \sqrt{2}d s_{32} =
%s_{22} A_{2}\; ,
%\nonumber
%\\
%\sqrt{2} d s_{22} + 2d^{2} s_{32} = s_{32} A_{2}\; ;
%\nonumber
%\\[3mm]
%2c^{2} s_{13} + \sqrt{2} c s_{23}  = s_{13} A_{3}\; ,
%\nonumber
%\\
%\sqrt{2} c  s_{13} +(c^{2}+d^{2} +1) s_{23} + \sqrt{2}d s_{33} =
%s_{23} A_{3}\; ,
%\nonumber
%\\
%\sqrt{2} d s_{23} + 2d^{2} s_{33} = s_{33} A_{3}\; . \label{4.1}
%\end{eqnarray}

With $A$ being $A_{1,2,3},$ the equations (\ref{3.3a})-(\ref{3.3c}) become linearly dependent in each subset of three equations. Consequently, one may drop an equation in each of the systems. Let us omit the third equation and put $s_{11}=s_{22} =s_{33}=1$. This results in the relations
\begin{eqnarray}
 s_{21} = - { (2c^{2} -A_{1}) \over \sqrt{2} c}\; ,\quad
 s_{31}=   {d   \over (2d^{2} - A_{1})}  { (2c^{2} -A_{1}) \over  c} \; ,
\label{4.2.1}
\\
 s_{12} = - {\sqrt{2} c \over (2c^{2} -A_{2})} \; ,\quad
 s_{32}= - { \sqrt{2} d  \over (2d^{2} -A_{2})} \; ,
\label{4.2.2}
\\
 s_{13} = { c \over (2c^{2} -A_{3})   }{ (2d^{2} -A_{3})  \over  d} \;,
\quad  s_{23} =  -{ (2d^{2} -A_{3})  \over \sqrt{2} d} \; .
\label{4.2.3}
\end{eqnarray}

\noindent Thus, the transformation
\begin{eqnarray}
S \Psi' = \left | \begin{array}{ccc}
1 & s_{12} & s_{13} \\
s_{21} & 1 & s_{23} \\
s_{31} & s_{32} & 1
\end{array} \right |
\left  | \begin{array}{c}
\Psi_{1}' \\
\Psi_{2}' \\
\Psi_{3}'
\end{array} \right |
  =
\left  | \begin{array}{c}
\Psi_{1} \\
\Psi_{2} \\
\Psi_{3}
\end{array} \right |
\label{4.3a}
\end{eqnarray}

\noindent reduces the radial system to the diagonal form:
\begin{eqnarray}
 \bar{\Delta}
\left|  \begin{array}{c}
\Psi_{1}' \\
\Psi'_{2} \\
\Psi'_{3} \
\end{array} \right |
  =
\left | \begin{array}{ccc}
A_{1} & 0 & 0\\
0 & A_{2} & 0\\
0 &0 & A_{3}
\end{array}  \right |
\left  | \begin{array}{c}
\Psi_{1}' \\
\Psi'_{2} \\
\Psi'_{3}
\end{array} \right |.
\label{4.3b}
\end{eqnarray}
Since the equations here are not connected, one can study three linearly independent solutions given as
\begin{eqnarray}
\left | \begin{array}{c}
\Psi'_{1} \\
0\\
0
\end{array} \right |, \qquad
\left | \begin{array}{c}
0 \\
\Psi'_{2}\\
0
\end{array} \right |, \qquad
\left | \begin{array}{c}
0 \\
0\\
\Psi'_{3}
\end{array} \right | .
\label{4.3c}
\end{eqnarray}

Thus, the problem is reduced to the second order differential equation
\begin{eqnarray}
\left (  {d ^{2}\over dr^{2} } + {2 \over r} {d \over dr}  + 2E  M
-{ L(L+1) \over r^{2} } \right ) f(r) = 0\; ,
\label{4.4}
\end{eqnarray}
where
\begin{eqnarray}
L(L+1) =2A=  \{ 2A_{1} ,\;  2A_{2} , \;  2A_{3} \}\, \quad \Rightarrow \quad L =
-{1 \over 2} \pm \sqrt{{1 \over 4} + 2A}\; .
\label{4.4b}
\end{eqnarray}

\noindent Note that the positiveness of the roots $A_{1,2,3}$ ensures the existence of positive values for parameter $L$ (upper sign in (\ref{4.4b})).

The above formalism is readily generalized to take into account the Coulomb potential. This is achieved by the change $E \rightarrow E+{\alpha/r}$ in the differential equation (\ref{4.4}):
\begin{eqnarray}
\left (  {d ^{2}\over dr^{2} } + {2 \over r} {d \over dr}  + 2
M \left(E  + {\alpha \over r} \right)  - { L(L+1) \over r^{2} } \right ) f(r) = 0\; .
\label{6.1a}
\end{eqnarray}
The structure of the singularities of this equation is the same as that for Eq. (\ref{4.4}): it has a regular singularity at $r=0$ and an irregular one of rank 1 at $r=\infty$ . Hence, the standard substitution with a scaled independent variable \cite{Slavyanov-Lay-2000}
\begin{eqnarray}
f = z^{L} e^{-z/2}F(z), \qquad z = 2 \sqrt{-2ME}\; r, \qquad -{\alpha \, \sqrt{-2ME} \over E }=B> 0 \; ,
\end{eqnarray}
reduces this equation to the confluent hypergeometric equation \cite{Abramowitz-Stegun}
\begin{eqnarray}
{d^{2}F  \over dz^{2}}  +  \left( {2(L+1)\over z}  -  1 \right) {d F \over dz } - {L+1-{B/2} \over z} F = 0\, ,
%\eqno(6.4)
\end{eqnarray}
the solution of which, restricted at the origin, is the Kummer confluent hypergeometric function: $F(z)={_1F_1}(a;b;z)$ with $a=L+1-{B/2}$ and $b=2(L+1)$. Applying now the condition for the Kummer function to turn into a polynomial, $ a = - n, \; n=0,1,2, ...$, we get the quantization rule for energy levels. In accordance with the three possible values for  $L$, we get three series of energy levels:
\begin{eqnarray}
E_{i} = - {1 \over 2 }\; {\alpha^{2}\,M \over (n+L_{i}+1)^{2} } ,\; \qquad
L _{i}  = -{1 \over 2} +  \sqrt{{1 \over 4} + 2A_{i}}\; .
\label{6.6b}
\end{eqnarray}
For the minimum value of $j$ the corresponding result is (see (\ref{5.4}):
\begin{eqnarray}
\left (  {d ^{2} \over dr^{2} } + {2 \over r} {d \over dr}  + 2M
\left(E + { \alpha \over r} \right)   \right ) \Psi_{1}= 0 \; \quad
\Rightarrow \quad
E = - {1 \over 2 }\; {\alpha^{2}\,M \over (n +1)^{2} } \; . \label{6.7a}
\end{eqnarray}

The case of the oscillator potential is treated in a similar way. This time, we have the differential equation
\begin{eqnarray}
\left [    {d^{2} \over dr^{2}}  + {2 \over r}{d \over dr}   + 2M
\left(E -  {\, kr^{2} \over 2} \right)
  - {L (L +1) \over  r^{2}}       \right ] f = 0 \; .
\label{7.1}
\end{eqnarray}

\noindent  Applying the substitution $ f(x)  = x^{L/2} e^{-x/2}\; F (x)$, with the new variable $x= \sqrt{ Mk}\,r^{2}$, we again arrive at the confluent hypergeometric equation having the solution $F(z)={_1F_1}(a;b;z)$ with the parameters
\begin{eqnarray}
a= {1 \over 2} \left ( {3\over 2}+ L - E    \sqrt{{M\over k}}
\right )  , \qquad b= L +{3 \over 2} \; .
\end{eqnarray}

\noindent The quantization rule $ a = -n$ again leads to three
series for the energy levels:
\begin{eqnarray} E_{i}  = {1\over 2}\,
\sqrt{{k \over M}} \,  \left({3 \over 2} + L_{i}+ 2n \right) \; , \qquad L_{i}
= -{1 \over 2} \pm \sqrt{{1 \over 4} + 2A_{i}} \; . \label{7.4}
\end{eqnarray}
This time, for the minimum value of $j$ we have
\begin{eqnarray}
\left [  {d ^{2} \over dr^{2} } + {2 \over r} {d \over dr}  + 2M
\left (E - {kr^{2} \over 2} \right  ) \right ] \Psi_{1}= 0 \;,
\quad E = {1\over 2}\, \sqrt{{k \over M}} \,  ({3 \over 2} + 2n) .
\label{7.5a}
\end{eqnarray}

More complicated equations arise if one considers other potentials. For instance, in the simultaneous presence of Coulomb, oscillator and linear potentials we have the equation
\begin{eqnarray}
  \left [   { d^{2}
\over d  r^{2} } +{2 \over r} {d  \over d r}   - {L(L+1) \over
r^{2} }
 +  {2M \over \hbar^{2}} \left (  E + {\alpha \over r}  - \beta r - kr^{2} \right )
  \right ]  R  = 0 \;
\label{Eq-1}
\end{eqnarray}
describing the quarkonium in particle physics, a flavorless meson whose constituents
 are a quark and its antiquark, that has been a subject of intensive discussions
 for already more than a half century (see, e.g., \cite{1}
 and references therein; for a recent review, see, e.g.,
 \cite{Quarkonium} and references therein). This equation is reduced to the
 biconfluent Heun equation \cite{Ovsiyuk-Veko-Amirf}, which is one of the
 natural polynomial generalizations of the confluent
  hypergeometric equation \cite{Ronveaux-1995, Slavyanov-Lay-2000}.

\section{ Spin 1 particle in the Lobachevsky space}

We now extend the above approach to the Lobachevsky geometry background. In spherical coordinates and tetrad
\begin{eqnarray}
dS^{2}= c^{2} dt^{2}-d r^{2}- \mbox{sh}^{2} r \; (d\theta^{2}+
\mbox{sh}^{2}{\theta}d\phi^{2})\,, \label{Lob}
\end{eqnarray}
\begin{eqnarray}
e^{\alpha}_{(0)}=(1, 0, 0, 0) ,
\quad e^{\alpha}_{(1)}=\left(0, 0, \frac {1}{\mbox{sh}\; r}, 0 \right) ,
\quad e^{\alpha}_{(2)}=\left(1, 0, 0, \frac{1}{ \mbox{sh}\; r  \sin \theta} \right) ,
\quad e^{\alpha}_{(3)}=(0, 1, 0, 0)
\end{eqnarray}

\noindent the Duffin-Kemer-Petiau equation reads (see the notations in
\cite{Book-1, Book-2})
\begin{eqnarray}
\hat{W} _{0}(x) \Phi^{0} (x)=\left [ i\;\beta ^{0}  \partial _{t} \; + \; i \left ( \beta ^{3} \;
\partial _{r}\; + {1 \over \mbox{tanh}\; r } \;(\beta ^{1} \; j^{31}  + \beta
^{2} \; j^{32}) \right ) +  {1 \over \mbox{sh}\;r}   \Sigma
^{k}_{\theta ,\phi } \; - \; M  \right ]    \Phi^{0} (x)  = 0 \; .
 \label{8.1}
\end{eqnarray}
Here the angular operator $\Sigma ^{k}_{\theta ,\phi }$ is defined as
\begin{eqnarray}
\Sigma ^{k}_{\theta ,\phi } \; = \;
 i\; \beta ^{1} \; \partial _{\theta } \; + \;
\beta ^{2} \; {i\; \partial_{\phi} \; + \; (i\; j^{12} - k) \cos
\theta \over \sin \theta} \; ,
\end{eqnarray}

\noindent where $\beta^{a}$ stand for $(10\times 10)$-matrices of Duffin-Kemer-Petiau equation,
$j^{ab} = \beta ^{a} \beta^{b}- \beta^{b}\beta^{a}$,  and the parameter $k = eg / \hbar c$ is quantized according to the Dirac rule:  $\mid k \mid  =1/2, \,1,\, 3/2,\, 2,\,  ...$ ($e$ and $g$ are electric and magnetic charges, respectively) \cite{1931-Dirac,1948-Dirac}.

The Duffin-Kemer-Petiau equation (\ref{8.1}) suggests a difficult problem, the treatment of which is rather complicated even in the nonrelativistic approximation. However, it turns out that a complete analytical study can be successfully performed if one consideres a slightly modified equation:
\begin{eqnarray}
\left [ i\;\beta ^{0}  \partial _{t} \; + \; i \left ( \beta ^{3}
\;
\partial _{r}\; + {1 \over \mbox{sinh}\; r } \;(\beta ^{1} \; j^{31}  + \beta
^{2} \; j^{32}) \right ) +  {1 \over \mbox{sh}\;r}   \Sigma
^{k}_{\theta ,\phi } \; - \; M  \right ]    \Phi (x)  = 0   \;.
   \label{8.2}
   \end{eqnarray}
The transition from (\ref{8.1}) to (\ref{8.2}) is achieved by introduction of an additional interaction term:
\begin{eqnarray}
\left [\hat{W}_{0}(x) + \Delta (r)  \right ] \Phi (x) =0 \; .
%\qquad \hat{W}_{0} \Longrightarrow \hat{W} (x)\;, \quad \Phi_{0}
%(x) \Longrightarrow \;\;\Phi (x) \; ,
\label{8.3a}
\end{eqnarray}

\noindent with
\begin{eqnarray}
\Delta (r) = i {1 - \cosh r \over \sinh r} \; (\beta^{1} J^{31} +
\beta^{2} J^{32} ) \; . \label{8.3b}
\end{eqnarray}
In the Cartesian tetrad gauge this operator is written as
 \begin{eqnarray}
\Delta^{Cart} (x)  = {1 - \cosh r \over \sinh r}  \;\left [
\vec{n} \cdot ( \vec{\beta} \times \vec{S} ) \right ]
 \; .
\label{8.10b}
\end{eqnarray}

%
%\noindent This term can be interpreted as an additional special interaction of spin 1 particle with the background geometry of the Lobachevsky space.

 \section{Separation of the variables in the presence of a magnetic monopole}

Consider the modified Duffin-Kemer-Petiau equation (\ref{8.2}) in spherical coordinates and tetrad. The components of the total
angular  momentum in this basis are determined by the formulas
\begin{eqnarray}
J^{k}_{1} =  l_{1} + { \cos \phi \over \sin \theta} \;(iJ^{12} -
\kappa)  \; , \qquad J^{k}_{2} = \;l_{2} + { \sin \phi \over \sin
\theta} \; (iJ^{12} - \kappa) \;  , \qquad J^{k}_{3} = l_{3} \;  .
\label{1.2a}
\end{eqnarray}

\noindent The wave functions of a particle with quantum numbers $(\epsilon , \;j ,\; m )$ are constructed using the substitution \cite{Book-2}
\begin{eqnarray}
\Phi _{\epsilon jm}(x)  = e^{-i\epsilon t} \; [\; f_{1}(r)\; D_{k}
,\;  f_{2}(r)\; D_{k-1} ,
 \; f_{3}(r) \; D_{\kappa},
\; f_{4}(r) \; D_{k+1}, \nonumber
\\
 f_{5}(r) \; D_{k-1}, \;  f_{6}(r) \; D_{k} , \;
 f_{7}(r) \; D_{k+1}, \;
f_{8}(r) \;D_{k-1} , \; f_{9}(r) \;D_{k} , \; f_{10}(r)\; D_{k+1}
\; ]  \;, \label{1.2b}
\end{eqnarray}

\noindent where the symbol $D_{\sigma}$ stands for the Wigner functions $D^{j}_{-m, \sigma}  (\phi, \theta, 0)$ \cite{VMH}. The corresponding radial equations read (note that $c = {1 \over 2}  \sqrt{(j + k)(j - k+ 1)} $ )
\begin{eqnarray}
-\left( {d \over dr} + {2 \over \mbox{sh}\; r\; } \right) \; f_{6} -
{\sqrt{2} \over \mbox{sh}\; r\; } \; \left( c \; f_{5} + d \; f_{7} \right)  -
M \; f_{1} = 0 \; , \nonumber
\\
i \epsilon \;  f_{5} + i \left( {d \over dr} + {1 \over \mbox{sh}\; r\;
\;} \right) \; f_{8} + i {\sqrt{2} c \over \mbox{sh}\; r\; \;}\; f_{9} -
M \; f_{2} = 0 \; , \nonumber
\\
i \epsilon \; f_{6} + {\sqrt{2} i \over \mbox{sh}\; r\; } \; (- c
\; f_{8} + d \; f_{10})
 - M \; f_{3} = 0 \; ,
\nonumber
\\
i \epsilon \; f_{7} - i \left( {d \over dr} + {1 \over \mbox{sh}\; r\;
\;} \right) \; f_{10} - i {\sqrt{2} d \over \mbox{sh}\; r\; \;}\;
 f_{9}  - M \; f_{4}= 0 \; ,
\nonumber
\end{eqnarray}
\begin{eqnarray}
-i \epsilon \; f_{2} +  {\sqrt{2} c \over \mbox{sh}\; r\; \;} \;
f_{1}  - M  \; f_{5} = 0 \; , \quad
 -i \epsilon \; f_{3}  - {d \over
dr} \; f_{1} - M \; f_{6} = 0 \; , \nonumber
\\
-i \epsilon \; f_{4} +{\sqrt{2} d \over \mbox{sh}\; r\; \;} \;
f_{1} - M \; f_{7} = 0 \; , \quad
 -i \left({d \over dr} + {1 \over \mbox{sh}\; r\; \;} \right) \; f_{2} -
i {\sqrt{2} c \over \mbox{sh}\; r\; \;} \;  f_{3} - M \; f_{8} = 0
\; , \nonumber
\\
i{\sqrt{2} \over \mbox{sh}\; r\; } ( c \; f_{2}  - d \; f_{4})  -
M \; f_{9} = 0 \; , \quad
i \left({d \over dr} + {1 \over \mbox{sh}\; r\; \;} \right) \; f_{4} + {i
\sqrt{2} d \over \mbox{sh}\; r\; \;} \; f_{3}  - M \; f_{10} = 0
\; . \label{38}
\end{eqnarray}

\noindent The quantum number $j$ adopts the values
\begin{eqnarray}
k = \pm  1/2 \; , \qquad  j = \mid k\mid  \; ,
 \mid k\mid  + 1 , \; \ldots \; ,
\end{eqnarray}
\begin{eqnarray}
k= \pm  1, \; \pm  3/2,\; \ldots \; , \qquad
 j = \mid k\mid - 1,\; \mid k\mid
,\; \mid k\mid  + 1, \; \ldots \; .
\end{eqnarray}

The cases for which $j = \mid k\mid - 1$ should be treated separately. For $k = + 1$ and $j = 0$, the appropriate substitution is
\begin{eqnarray}
\Phi^{(0)} (t,r) = \; e^{-i\epsilon t} \; ( \; 0 , \;\; \; f_{2}
,\;\;  0 , \;\;  0 , \;\; f_{5}, \;\; 0 ,\; 0 ; \;\; f_{8} , \;\;
0 , \;\; 0 \; ) \; .
\end{eqnarray}
This leads to the following three radial equations:
\begin{eqnarray}
f_{5} = \; - \; i \; {\epsilon \over M} \; f_{2} \; , \qquad f_{8}
= \; -\;  {i \over M} \; \left ( \; {d \over dr} \; + \; {1 \over
\mbox{sh}\; r\;} \right ) \;  f_{2}\; , \nonumber
\\
\left ( {d^{2} \over dr^{2}} + {2 \over \mbox{sh}\; r} {d \over
dr} + {1 - \mbox{ch}\; r \over  \mbox{sh}^{2} r } + \epsilon^{2} -
M^{2} \right ) f_{2} = 0 \; . \label{29}
\end{eqnarray}

\noindent By a further substitution, we simplify the problem as follows:
\begin{eqnarray}
f_{2}(r)  = {1 + \mbox{ch}\; r \over  2\; \mbox{sh}\; r}
F_{2}(r)\;, \qquad \left (\; {d^{2} \over dr^{2}} \; + \; \epsilon
^{2} \;   - \; M^{2} \; \right ) \; F_{2}  = 0 \; . \label{40}
\end{eqnarray}

\noindent The last equation coincides with the one appearing in the flat space \cite{Book-2}. The case $j= 0$, $k = - 1$ yields the same result.

\section{Nonrelativistic approximation in the general case}

As in the previous Minkowski case, to derive a non-relativistic approximation, we first exclude from the radial equations (\ref{38}) the non-dynamical variables $f_{1}, f_{8}, f_{9},  f_{10}$. We then get a system, which, using the  more symmetric notations
\begin{eqnarray}
(f_{2}, f_{3}, f_{4} )  \rightarrow  (\Phi_{1},  \Phi_{2}, \Phi_{3})  \;
, \qquad (f_{5}, f_{6}, f_{7} ) \rightarrow   (E_{1}, E_{2}, E_{3}) \; ,
\nonumber
\end{eqnarray}

\noindent is written as
\begin{eqnarray}
i \epsilon M  E_{1}  + i \left ( {d \over dr} + {1 \over
\mbox{sh}\; r} \right ) \left [ -i \left ({d \over dr} + {1 \over
\mbox{sh}\; r} \right ) \Phi_{1} - i {\sqrt{2} c \over \mbox{sh}\;
r\;} \Phi_{2} \right ] +i {\sqrt{2} c \over \mbox{sh}\; r} \left
[i{\sqrt{2} \over r} ( c  \Phi_{1}  - d \Phi_{3}) \right  ] -
M^{2}  \Phi_{1}  = 0  , \nonumber
\\
i \epsilon M E_{2} + {\sqrt{2}i \over \mbox{sh}\; r} \left  [- c
 \left ( -i({d \over dr} + {1 \over \mbox{sh}\; r})  \Phi_{1} - i {\sqrt{2}
  c \over \mbox{sh}\; r}   \Phi_{2}  \right )
+ d  \left ( i ({d \over dr} + {1 \over \mbox{sh}\; r}) \Phi_{3} +
{i \sqrt{2} d \over \mbox{sh}\; r}  \Phi_{2}  \right ) \right ]
 - M^{2}  \Phi_{2} = 0  ,
\nonumber
\\
i \epsilon M E_{3} - i \left ( {d \over dr} + {1 \over \mbox{sh}\;
r  } \right ) \left [ i \left ({d \over dr} + {1 \over \mbox{sh}\;
r} \right )  \Phi_{3} + {i \sqrt{2} d \over \mbox{sh}\; r}
\Phi_{2} \right  ]  - i {\sqrt{2} d \over \mbox{sh}\; r\;}
 \left [i{\sqrt{2} \over \mbox{sh}\; r} ( c  \Phi_{1}  - d  \Phi_{3}) \right ]  - M^{2} \Phi_{3}= 0  ,
\nonumber
\end{eqnarray}
\begin{eqnarray}
-i \epsilon M  \Phi_{1} +  {\sqrt{2} c \over \mbox{sh}\; r\;}
\left [ - \left ( {d \over dr} + {2 \over \mbox{sh}\; r\;} \right  )  E_{2} -
{\sqrt{2} \over \mbox{sh}\; r\;}  ( c \; E_{1} + d  E_{3}) \right
]  - M^{2}  E_{1} = 0 \; ,
\nonumber
\\
-i \epsilon M \Phi_{2}  - {d \over dr}  \left [ - \left ( {d \over dr} +
{2 \over \mbox{sh}\; r\;}\right  )  E_{2} - {\sqrt{2} \over \mbox{sh}\;
r\;}  ( c  E_{1} + d \; E_{3} ) \right ] - M ^{2} E_{2} = 0 \; ,
\nonumber
\\
-i \epsilon M  \Phi_{3} +{\sqrt{2} d \over \mbox{sh}\; r\;} \left
[ - \left ( {d \over dr} + {2 \over \mbox{sh}\; r\;} \right )  E_{2} - {\sqrt{2}
\over \mbox{sh}\; r\;}  ( c  E_{1} + d
 E_{3} ) \right ] - M^{2}  E_{3} = 0  \; .
\label{2.3.5}
\end{eqnarray}
The big  $\Psi_{k} $ and small $\psi_{k}  $  components are now
introduced through the linear combinations \cite{Book-2})
\begin{eqnarray}
\Phi_{k}  = {\Psi_{k} + \psi_{k} \over 2} , \qquad   i E_{k}  =
{\Psi_{k} - \psi_{k} \over 2}
 \;. \label{2.3.6}
 \end{eqnarray}

\noindent Performing essentially the same calculations as in the case of the Minkowski space, we arrive at the following
nonrelativistic system for three big components  (note that  $\Psi_{i} ={1 + \mbox{ch}\, r \over \mbox{sn}\, r} F_{i}$
):
\begin{eqnarray}
\left (
 {d ^{2} \over dr^{2} }
   + 2E  M  -{4c^{2} \over \mbox{sh}^{2}\; r\; } \right )    F_{1} =
   {1 + \mbox{ch}\; r\; \over \mbox{sh}^{2} r }  \left (  \qquad    \sqrt{2} c\;
   F_{2} \qquad  \right ) ,
\nonumber
\\
 \left ( {d ^{2}\over dr^{2}} +
2EM  -{2 (c^{2} + d^{2} ) \over  \mbox{sh}^{2} r}  \right ) F_{2}
=
  {(1 + \mbox{ch}\; r )\over \mbox{sh}^{2} r }
  \left (  \sqrt{2} c  \;  F _{1} +  F_{2}
 +
\sqrt{2} d  \; F _{3}  \right ),
\nonumber
\\
\left (
 {d ^{2} \over dr^{2} } +  2E  M  -{4d^{2} \over \mbox{sh}^{2}\; r\; } \right )    F_{3} =
 {1 + \mbox{ch}\; r\; \over \mbox{sh}^{2} r }   \left ( \qquad  \sqrt{2} d
   \; F_{2} \qquad  \right ) .
\label{2.3.14}
\end{eqnarray}

This system is still very difficult to solve; the method used in the casee of the flat space cannot be applied here. However, we are able to solve the problem in the case of minimum value of $j= \mid k \mid -1$ in the presence of additional Coulomb or oscillator potentials.

\section{ Minimum value of $j$: the Coulomb and oscillator potentials }

Consider first the Coulomb  potential. The corresponding equation for this case reads \cite{Red'kov-Ovsiyuk-2012}
\begin{eqnarray}
\left (\; {d^{2} \over dr^{2}} \; + \; \left( \epsilon + { \alpha \over
\mbox{th}\; r} \right)^{2} \;   - \; M^{2} \; \right) F_{2}  = 0 \;.
\label{42}
\end{eqnarray}

\noindent Applying the substitution $ F_{2} = x ^{A}(1-x)^{B} f (x)$,  $x = 1 - e^{-2r}$, at positive
\begin{eqnarray}
A = {  1  + \sqrt{  1-4 \alpha^{2}} \over 2}  \; , \qquad  B=+{1
\over 2} \sqrt{ -(  \epsilon+  \alpha  )^{2}+   M^{2}   } \; ,
\end{eqnarray}
we get the hypergeometric equation:
\begin{eqnarray}
 x (1-x) f'' +   [  2A   - (2A + 2B +1) x  ] \; f'
   -  \left [ (A+B)^{2}
   + \left({  \epsilon \over 2}-{ \alpha  \over 2}\right)^{2}- {  M^{2}  \over 4}       \right ]  f  = 0.
\end{eqnarray}
Hence, $f(x)={_2F_1}(\lambda, \beta; \gamma;x)$, and the quantization rule  $\lambda = - n$ gives the energy spectrum
\begin{eqnarray}
\epsilon = { M \over \sqrt{1 + \alpha^{2} / \nu^{2}  }}\; \sqrt {
1 - {\alpha^{2} + \nu^{2} \over M^{2}}}, \qquad \nu = n + {1+
\sqrt{  1-4 \alpha^{2}}  \over 2} . \label{43}
\end{eqnarray}
In usual units this reads
\begin{eqnarray}
E = { mc^{2} \over \sqrt{1 + \alpha^{2} / \nu^{2}  }}\; \sqrt { 1
- {\hbar^{2} \over m^{2} c^{2} R^{2} }  (\alpha^{2} + \nu^{2})} \; ,
\label{44}
\end{eqnarray}
where $R$ stands for a curvature radius of the Lobachevsky space. We note that the number of bound states is finite.

In the case of the oscillator potential the corresponding equation \cite{Red'kov-Ovsiyuk-2012}:
\begin{eqnarray}
\left (\; {d^{2} \over dr^{2}} \; + \; 2 M (E  - { K \;
\mbox{th}^{2} r  \over 2} ) \right ) \; F_{2}  = 0 \;, \label{46}
\end{eqnarray}
is again solved in terms of the hypergeometric functions. The resultant energy spectrum reads
\begin{eqnarray}
E = N \sqrt{ {K \over M} + ({1 \over 2M})^{2}} -{1 \over 2M}
(N^{2} + {1 \over 4}) \;, \quad N = 2n +{3\over 2} \; \label{47}
\end{eqnarray}
or, in usual units,
\begin{eqnarray}
\epsilon = \hbar \left ( N \; \sqrt{ {k\over m} +  {\hbar^{2}
\over 4m^{2}R^{4} }} - {\hbar \over 2mR^{2}} \left(N^{2}+{1\over 4} \right) \right )  . \label{48}
\end{eqnarray}

\section{Spin 1 free particle in the absence of the monopole background}

In the absence of the monopole, the identity  $ d =c ={1 \over 2}
\sqrt{j(j+1)} $  holds, and Eqs. (\ref{2.3.14}) become simpler:
\begin{eqnarray}
\left (
 {d ^{2} \over dr^{2} }
   + 2E  M \right )    F_{1} - \sqrt{2} c
   {1 + \mbox{ch}\; r\; \over \mbox{sh}^{2} r }  F_{2}
-{4c^{2} \over \mbox{sh}^{2}\; r\; }   F_{1}  = 0 \; , \qquad  \qquad \qquad \nonumber
\\
\left (
 {d ^{2} \over dr^{2} } +  2E  M \right )    F_{3} - \sqrt{2} c
   {1 + \mbox{ch}\; r\; \over \mbox{sh}^{2} r }   F_{2}
-{4c^{2} \over \mbox{sh}^{2}\; r\; }  F_{3}  = 0 \; , \qquad \qquad  \qquad \nonumber
\\
 \left ( {d ^{2}\over dr^{2}} +
2EM  \right ) F_{2}  - {4 c^{2} \over  \mbox{sh}^{2} r}  F_{2}
 - {(1 + \mbox{ch}\; r )\over \mbox{sh}^{2} r }  F_{2}
- {\sqrt{2} c (1 + \mbox{ch}\; r )\over \mbox{sh}^{2} r }   F _{1}
- {\sqrt{2} c (1 + \mbox{ch}\; r )\over \mbox{sh}^{2} r }  F _{3}
=0 \; . \label{50}
\end{eqnarray}

\noindent Now, in contrary to the monopole case, one can additionally diagonalize the space reflection operator:
\begin{eqnarray}
P = (-1)^{j+1}, \;F_{2} = 0, \; F_{3} = - F_{1} \; ;
\quad\quad
P = (-1)^{j}, \;   F_{3} = + F_{1} \; . \label{51}
\end{eqnarray}
Correspondingly, the system (\ref{50}) is split into $1+2$ equations, namely,
\begin{eqnarray}
\left (
 {d ^{2} \over dr^{2} }    + 2E  M
  -{4c^{2} \over \mbox{sh}^{2}\; r\; }  \right )   F_{1}  = 0 \;
\label{52}
\end{eqnarray}
and ($\nu = c \sqrt{2} $)
\begin{eqnarray}
\bar{\Delta} \left | \begin{array}{c}
F_{1} \\
F_{2}
\end{array} \right | =
\left | \begin{array}{cc}
0 & \nu \\
2 \nu &  1
\end{array} \right |
\left | \begin{array}{c}
F_{1} \\
F_{2}
\end{array} \right | ,
\quad
 \bar{\Delta}= {\mbox{sh}^{2} r  \over  1 + \mbox{ch}\; r  }
\left (
 {d ^{2} \over dr^{2} }  -{j(j+1) \over \mbox{sh}^{2}\; r\; }     + 2E  M   \right ).
\label{53} \end{eqnarray}
For the last system (\ref{53}), one can diagonalize the mixing matrix:
\begin{eqnarray}
F = S F' \; , \qquad  \bar{\Delta} \;F' = S^{-1}A\;S F'\; , \qquad
S = \left | \begin{array}{cc}
s_{11} & s_{12} \\
s_{21} & s_{22}
\end{array} \right |,\;
\bar{\Delta}\; F' = \left | \begin{array}{cc}
j+1 & 0 \\
0 & -j
\end{array} \right | F' \; ,
\label{54}
\end{eqnarray}

\noindent so we get two disconnected equations:
\begin{eqnarray}
 \left (
 {d ^{2} \over dr^{2} } + 2E  M   -{j(j+1) \over \mbox{sh}^{2}\; r\; }
  - {1 + \mbox{ch}\; r  \over \mbox{sh}^{2}r }(j+1)     \right )    F'_{1}=0 \; ,
\label{55.a}
\\
\left ( {d ^{2}\over dr^{2}}  + 2E  M  -{j(j+1) \over
\mbox{sh}^{2} r}  +
 {1 + \mbox{ch}\; r  \over \mbox{sh}^{2} r }j   \right ) F'_{2}  = 0 \; ,
\label{55}
\end{eqnarray}

\noindent
which are of the same type. Using the variable
$ y = (\mbox{ch}\; r + 1)/2  $, we have
\begin{eqnarray}
 \left (
 y(y-1) {d^{2} \over dy^{2}} +  \left(y -{1 \over 2} \right) {d \over dy} + 2ME     -{j(j+1) \over 4 y(y-1) }
  + { \mu  \over 2 (y-1) }    \right )    F =0 \; ,
\label{56}
\end{eqnarray}
where $ \mu = -j-1, \;  + j\;$. The solution of this equation is written in terms of the hypergeometric function:
\begin{eqnarray}
F = y^{a} (1-y) ^{b} {_2F_1} ( \lambda, \beta; \gamma ; y) \; , \quad
\gamma = 2a +{1\over 2} \;, \;
\lambda = a+b  + i\sqrt{ 2ME} \;, \;  \beta = a+b  - i \sqrt{
2ME} \; , \label{57}
\end{eqnarray}
where
\begin{eqnarray}
a = {j+1 \over 2}, \; a'= -{j \over 2}\;;
\nonumber
\\
\mu=0\; , \qquad b ={j+1 \over 2}, \; b' = -{j \over 2}\; ;
\nonumber
\\
\mu =-j-1\; ,  \qquad b = {j+2\over 2} , \; b' = - {j+1 \over 2}
\nonumber
\; ;
\\
\mu =+j\; ,  \qquad b =  {j\over2},\; b' =  - { j+1  \over 2}  \; .
\end{eqnarray}

\noindent Since $y=1$ corresponds to the point $r=0$, in order to have finite solutions at the origin, we should take $b>0$. The asymptotic behavior of the solution at infinity then is
\begin{eqnarray}
f(y) \sim {\Gamma(\gamma) \Gamma(\beta - \alpha) \over \Gamma
(\gamma - \alpha) \Gamma(\beta)}
 \left(- { e^{r} \over 4} \right)^{-i\sqrt{2EM}}
 + {\Gamma(\gamma) \Gamma( \alpha - \beta)  \over \Gamma (\gamma - \beta) \Gamma(\alpha)}
        \left(-{ e^{r} \over 4} \right) ^{+ i\sqrt{2EM} }\; .
\label{58}
\end{eqnarray}
Thus, we have derived real solutions, regular at $r=0$ and having standing-wave behavior at infinity.

\section{ Spin 1 particle in the Coulomb attractive potential}

In the presence of the external Coulomb field, instead of equations (\ref{52}), (\ref{55.a}) and (\ref{55}) we have

\begin{eqnarray}
\left (
 {d ^{2} \over dr^{2} }    + 2  M \left(E + {\alpha \over \mbox{tanh}\; r} \right)
  -{j(j+1) \over \mbox{sh}^{2}\; r\; }  \right )   F_{1}  = 0 \; ,
\label{59}
\\
 \left (
 {d ^{2} \over dr^{2} } + 2  M \left(E + {\alpha \over \mbox{tanh}\; r} \right)   -{j(j+1) \over \mbox{sh}^{2}\; r\; }
  - {1 + \mbox{ch}\; r  \over \mbox{sh}^{2} r}(j+1)     \right )    F'_{1}=0 \; ,
\label{60.1}
\\
\left ( {d ^{2}\over dr^{2}}  + 2  M \left(E + {\alpha \over
\mbox{tanh}\; r} \right) -{j(j+1) \over  \mbox{sh}^{2} r}  +
 {1 + \mbox{ch}\; r  \over \mbox{sh}^{2} r }j   \right ) F'_{2}  = 0 \; .
\label{60.2}
\end{eqnarray}

Passing to the new variable $F_{1} = x ^{a}(1-x)^{b} f (x)$ with $ x = 1 - e^{-2r}$, Eq. (\ref{59}) reads
%\begin{eqnarray}
%\left [
% (1-x)^{2}{d^{2} \over d x^{2}}- (1-x){d \over d x}
%    + {  M \over 2} \left(E + \alpha { 2-x \over  x} \right)
%  -  j(j+1){ 1-x  \over x^{2}  }  \right ]   F_{1}  = 0  \; .
%\label{5.2b}
%\end{eqnarray}
\begin{eqnarray}
 x (1-x) f'' +   [  2a   - (2a + 2b +1 ) x  ]\; f' +
\nonumber
\\
\left [ {a (a-1) -j(j+1)  \over x} + \left( b^{2} + {M \over 2} (E + \alpha) \right)   {1 \over 1-x}  -
   a ^{2}   - 2a b  - b^{2}
        - {  M  \over 2}  (E - \alpha)        \right ]  f  = 0  \; .
\label{5.2c}
\end{eqnarray}

\noindent Putting now $a = j+1, -j$ and $b =\pm   \sqrt{ -{M\over 2}(E+ \alpha)}$, we obtain the hypergeometric equation with the solution $f(x)={_2F_1} ( \lambda, \beta; \gamma; x)$, for which in order to get bound states, we take
\begin{eqnarray}
\lambda = a+ b - \sqrt{ -{  M  \over 2}  (E - \alpha)}  \; ,
\qquad \beta = a+ b +  \sqrt{ -{  M  \over 2}  (E - \alpha)}
\end{eqnarray}
with
\begin{eqnarray}
a = j+1 \; , \qquad b = + \sqrt{ -{M\over 2} ( E +\alpha) } \; .
\end{eqnarray}
The hypergeometric function turns into polynomial if $\lambda = - n$. This gives the energy levels as
\begin{eqnarray}
E= -{M\alpha^{2} \over 2N^{2}} - {N^{2} \over 2M} \; , \qquad N =  j+1 + n \; .
\label{energy}
\end{eqnarray}

\noindent
In usual units this formula reads
\begin{eqnarray}
\epsilon = - mc^{2} {\alpha ^{2} \over 2 N^{2}} - {\hbar^{2} \over m R^{2} } {N^{2} \over 2} \qquad
\left (\alpha = {e^{2} \over \hbar c}  = {1 \over 137 }  \right )\; .
\label{energy'}
\end{eqnarray}

Now consider Eqs. (\ref{60.1}) and (\ref{60.2}). Using the variable $z = \mbox{th} (r  /  2)$, they are written as
\begin{eqnarray}
 \left [
 { d ^{2}\over dz^{2} } - {2 z \over 1- z^{2}}  {d \over dz} +
  8 M \left(E + \alpha  {1+ z^{2} \over 2z} \right) {1 \over (1-z^{2})^{2}}   -{j(j+1)  \over z^{2} }
  - {2(j+1)   \over z^{2}(1-z^{2})  }    \right ]    F'_{1}=0 \; ,
\label{64}
\end{eqnarray}
\begin{eqnarray}
\left [
  { d ^{2}\over dz^{2} } - {2z \over 1- z^{2}} {d \over dz}
+ 8  M \left(E + \alpha {1+ z^{2} \over 2z} \right) {1 \over (1-z^{2})^{2}}
-{j(j+1)   \over z^{2} }  +
  {2 j   \over z^{2} (1 -z^{2})  }   \right ] F'_{2}  = 0 \; .\label{65}
\end{eqnarray}

\noindent These equations have four regular singular points located at $z=0, \pm 1, \infty$, two of which are physical: $z = 0 \;(r =0), \; z = +1 \; (r = \infty)$. Applying the simplifying substitution
\begin{eqnarray}
F'_{1}(z)={1 \over {\sqrt{z^{2}-1}}} z^{A}(1-z)^{B}(-1-z)^{C}H(z)\,
\end{eqnarray}
with $A, \,B,\,C$ taken as
\begin{eqnarray}
 A=-(j+1),\;j+2\,,\quad
  B={1\over
2}\pm\sqrt{-2M(E+\alpha)}\,,\quad  C={1\over
2}\pm\sqrt{-2M(E-\alpha)}\,,
\end{eqnarray}

\noindent Eq. (\ref{64}) is reduced to the general Heun equation
\begin{eqnarray}
{d^{2}\, H  \over dz^{2}} +
\left ({\gamma \over z} +{\delta \over
z-1 } +{\epsilon \over z+1 }\right )\,{dH \over dz}+
 {\lambda \beta z - q \over z(z-1)(z+1) } H =0 , \quad ( \gamma  + \delta  + \epsilon = \lambda + \beta + 1)
\end{eqnarray}

\noindent  with parameters
\begin{eqnarray}
\gamma = 2A\; , \quad \delta = 2 B \; , \quad \epsilon = 2 C \; , \quad q=4M\alpha-2A(B-C)\; ,
\nonumber
\\
\lambda  = -j-1 +
(A+B+C ), \quad \beta  = j + (A+B+C)\; .
\end{eqnarray}

We now use a quantization condition of the form $\beta  = - n$. The appropriate choice for $A,B,C$ then is
\begin{eqnarray}
A=j+2\,, \qquad B={1\over 2} + \sqrt{-2M(E+\alpha)}\,, \qquad
C={1\over 2} - \sqrt{-2M(E-\alpha)}\,.
\end{eqnarray}
This leads to the following formula for the energy levels:
\begin{eqnarray}
E=-{M\alpha^{2}\over 2(j+ 3/2 +n/2 )^{2}} -{(j + 3/2
+n/2)^{2}\over 2M} \,. \label{70}
\end{eqnarray}

%-----------------------------------------------------------------------------------

%A more special situation happens when the parameters entering HeunG are
 %such that the function is, simultaneously, a Frobenius solution around three adjacent singularities and hence analytic
 %in a domain containing all of them. In such a case the solution will also be a Frobenius
 %solution around the fourth singularity and Heun  G will be a polynomial. A necessary
 %(not sufficient) condition for this case is that $\alpha = -n$, with $n$ a positive integer, and $q$
% has one of a finite number of characteristic values, in which case the function is a polynomial of degree $n$.

%-----------------------------------------------------------------------------------

\noindent
In the similar manner, Eq. (\ref{65}) gives the spectrum
\begin{eqnarray}
E= -{M\alpha^{2} \over 2(j+1/2 + n/2 )^{2}} - {(j+1/2 + n/2)^{2}
\over 2M} \;  . \label{71}
\end{eqnarray}

%Thus we have found three series of energy levels: (\ref{61}),
%(\ref{70}),  (\ref{71}). The presence of  $n' $ and   $n'/2$ is
%due to the use of different variables in solving respective
%differential equations,
% $z= \mbox{th}\; {r\over 2}$ and  $x =1-e^{-2r}$ related by quadratical relations:
%$$
%x = { 2 \mbox{th}\; r \over 1 + \mbox{th}\; r} ,\qquad \mbox{th}\;
%r = {2z \over 1 +z^{2}}, \qquad x = {4z (1+z^{2} )\over
%(1+z^{2})^{2} + 4z^{2} } ={4 (z+z^{-1}) \over 4 + (z+z^{-1)}} \;.
%$$

\section{Particle in the oscillator field}

In the presence of the oscillator potential, the radial equations take the form

\vspace{2mm}
$P = (-1)^{j+1}$:
\begin{eqnarray}
\left [
 {d ^{2} \over dr^{2} }    + 2  M \left (E - {K \; \mbox{th}^{2} r \over 2} \right )
  -{j(j+1) \over \mbox{sh}^{2}\; r\; }  \right ]   F_{1}  = 0 \; ,
\label{2.7.1a}
\end{eqnarray}
and
$P = (-1)^{j}$:
\begin{eqnarray}
 \left [
 {d ^{2} \over dr^{2} } + 2  M \left (E - {K \; \mbox{th}^{2} r \over 2}\right
  )   -{j(j+1) \over \mbox{sh}^{2}\; r\; }
  - {1 + \mbox{ch}\; r  \over \mbox{sh}^{2} r}(j+1)     \right ]    F'_{1}=0 \; ,
\label{2.7.1c}
\\
\left [ {d ^{2}\over dr^{2}}  + 2  M \left  (E - {K \; \mbox{th}^{2} r
\over 2} \right ) -{j(j+1) \over  \mbox{sh}^{2} r}  +
 {1 + \mbox{ch}\; r  \over \mbox{sh}^{2} r }j   \right ] F'_{2}  = 0
\; . \label{2.7.1c'}
\end{eqnarray}

\noindent By means of the transformation of the independent variable $ x = \mbox{ch}\; r $ the coefficients of the equations become rational functions:

\vspace{2mm}
$P = (-1)^{j+1},$
\begin{eqnarray}
\left  [ (x^{2}- 1) {d^{2} \over d x^{2}} + x {d \over d x}
   + 2  M \left (E - {K\over 2} {x^{2} -1  \over x^{2} } \right )
  -{j(j+1) \over x^{2} - 1 }  \right ]   F_{1}  = 0 \; ;
\label{2.7.3a}
\end{eqnarray}
and
$P = (-1)^{j},$
\begin{eqnarray}
 \left [
 (x^{2}- 1) {d^{2} \over d x^{2}} + x {d \over d x} +
  2  M  \left (E - {K\over 2} {x^{2} -1  \over x^{2} } \right )   -{j(j+1) \over x^{2} - 1 }
  - {1 + x  \over x^{2} - 1}(j+1)     \right ]    F'_{1}=0 \; ,
\label{2.7.3c}
\\
\left [ (x^{2}- 1) {d^{2} \over d x^{2}} + x {d \over d x}  +
 2  M \left (E - {K\over 2} {x^{2} -1  \over x^{2} } \right ) -{j(j+1) \over  x^{2} - 1 }  +
 {1 + x  \over x^{2} - 1 }j   \right ] F'_{2}  = 0
\; .
\label{2.7.3c'}
\end{eqnarray}

%These are again Fuchsian differential equations having for regular singular points: $z=0, \pm 1, \infty$, so that they can be reduced to the general Heun equation. For Eq. (\ref{2.7.3a}), however, the characteristic exponents of the singularities $z=+1$ and $z=-1$ are equal, hence, one can use the quadratic transformation of the  variable $y=x^{2}$ to derive a simpler equation of the hypergeometric type:

For Eq. (\ref{2.7.3a}), one can further apply the quadratic transformation of the  variable $y=x^{2}$ to derive an equation of the hypergeometric type:
\begin{eqnarray}
\left ( 4y(y-1)  {d^{2} \over d y^{2}} + (4y-  2 ) {d \over d y}
   + 2  M E - M K {y -1  \over y }
  -{j(j+1) \over y - 1 }  \right )   F_{1}  = 0 \; .
\label{2.7.4a}
\end{eqnarray}

\noindent By means of the substitution  $F_{1}(y)=y^{a} (1-y)^{b} f(y)$ with
\begin{eqnarray}
a={1 \over 4}\pm{1 \over 4}\sqrt{1+4KM},\qquad b=-{j \over 2}, \; {1+j
\over 2} \; , \label{2.7.4b}
\end{eqnarray}

\noindent this equation is reduced to the canonical form of the hypergeometric equation:
\begin{eqnarray}
x(1-x) {d^{2} \over dx^{2} } f + [ \gamma + (\lambda + \beta +1) x
] {d f \over d x} - \lambda \beta \; f =0 \;
\end{eqnarray}
with parameters
\begin{eqnarray}
\gamma =2a+1, \quad
\lambda = a+b -  \sqrt{ - {2EM \over 4} +{KM \over 4}}  \; , \quad
 \beta
 = a+b +  \sqrt{ - {2EM \over 4} + {KM \over 4}}  \; .
\end{eqnarray}
Accordingly, the solution of Eq. (\ref{2.7.3a}) is written as
\begin{eqnarray}
F_{1} (y) = y^{a} (1-y)^{b} {_2F_1}(\lambda, \beta, \gamma, y)
=(\mbox{ch}\; r)^{2a} (-\mbox{sh}\; r )^{2b} {_2F_1}(\lambda, \beta,
\gamma, \mbox{ch}^{2}r)\; . \label{2.7.5a}
\end{eqnarray}

To construct solutions associated with bound states, we should take
\begin{eqnarray}
2b = (1+j) > 0 \;, \qquad  2a = {1 - \sqrt{1 +4KM} \over 2} < 0 \; .
\label{2.7.5b}
\end{eqnarray}
Then, the condition to have polynomial solutions, $\lambda = -n$, leads to the result
\begin{eqnarray}
E = N \sqrt{ {K \over M} + \left({1 \over 2M} \right)^{2}} -{1 \over 2M}
\left(N^{2} + {1 \over 4} \right), \; \quad N = 2n +j +{3\over 2} . \label{2.7.6b}
\end{eqnarray}
We note that from the structure of the solution, Eq. (\ref{2.7.5a}), it is readily seen an inequality which ensures the vanishing of the solution at infinity:
\begin{eqnarray}
a+b + n <  0 \qquad \Longrightarrow \qquad { 1 - \sqrt{1+4KM}
\over 4}  + {j+1 \over 2} + n < 0 \; .
\label{2.7.7b}
\end{eqnarray}

\noindent This gives the upper limit for the number of the possible bound states:
\begin{eqnarray}
2n + j + {3\over 2} < { \sqrt{1+4KM} \over 2} \; .
\label{2.7.7c}
\end{eqnarray}
In  usual units, the spectrum is written as
\begin{eqnarray}
\epsilon = \hbar \left ( N \; \sqrt{ {k\over m} +  {\hbar^{2}
\over 4m^{2}R^{4} }} - {\hbar \over 2mR^{2}} \left(N^{2}+{1\over 4}\right) \right ), \qquad N = 2n +j +{3\over 2}\;  \label{2.7.8}
\end{eqnarray}
and the restriction imposed on the quantum numbers reads
\begin{eqnarray}
2n + j + {3\over 2} <  {1 \over 2}  \sqrt{1+ {4km \over \hbar^{2}}
R^{4} }  \; \; . \label{2.7.10}
\end{eqnarray}

Now, let us turn to Eq. (\ref{2.7.3c}). Applying the substitution
\begin{eqnarray}
F'_{1}=x^{A}(1-x)^{B}(-1-x)^{C}H(x)\, \label{2.7.11a}
\end{eqnarray}
with $A,\,B,\,C$ chosen as
\begin{eqnarray}
 A={1\over 2}\pm\,{1\over
2}\sqrt{1+4\,MK}\,,\quad B=-{1\over
2}-{j\over2}\,,\;\;1+{j\over 2}\,,\quad  C=-{j\over
2}\,,\;\;{1\over 2}+{j\over 2}\,,
 \label{2.7.11b}
 \end{eqnarray}

\noindent we arrive at the general Heun equation
\begin{eqnarray}
{d^{2}\, H  \over dz^{2}} +
\left ({\gamma \over z} +{\delta \over
z-1 } +{\epsilon \over z+1 }\right )\,{dH \over dz}+
 {\lambda \beta z - q \over z(z-1)(z+1) } H =0 , \quad ( \gamma  + \delta  + \epsilon = \lambda + \beta + 1)
\end{eqnarray}
with parameters
\begin{eqnarray}
\gamma = 2A\,,  \quad \delta  =2B+{1\over 2}\,, \quad \epsilon = 2  C +{1 \over 2} \,, \quad q=-2A(B-C)\,,
\\
\lambda  = A+B+C+\sqrt{-M(2E-K)}\; ,\quad
\beta  =  A+B+C -\sqrt{-M(2E-K)}\; . \label{2.7.11e}
\end{eqnarray}

\noindent As a formal quantization condition, we now apply one of the two necessary conditions for polynomial solutions: $ \beta =-n $ (we stress that since we use only one of the two conditions, we actually do not construct  polynomials). At
\begin{eqnarray}
A={1\over 2} - \,{1\over 2}\sqrt{1+4\,MK}\,,\qquad B=1+{j\over
2}\,, \qquad C={1\over 2}+{j\over 2}\,\label{2.7.12a}
\end{eqnarray}

\noindent  this  condition gives the energy spectrum
\begin{eqnarray}
E = N \sqrt{ {K \over M} + \left({1 \over 2M} \right)^{2}} -{1 \over 2M}
\left(N^{2} + {1 \over 4} \right) , \; \quad N= 2+j +n . \label{2.7.13b}
\end{eqnarray}

In the similar manner, Eq. (\ref{2.7.3c'}) leads to the same spectrum but with $N= 1+j +n$.

%Thus,  collecting the results together, we write down three series  for energy values:
%\begin{eqnarray}
%\epsilon = \hbar \left ( N \; \sqrt{ {k\over m} +  {\hbar^{2}
%\over 4m^{2}R^{4} }} -  {\hbar \over 2mR^{2}} (N^{2}+{1\over 4})
%\right ), \; N = 2n +j +{3\over 2}\; , \nonumber
%\\
%\epsilon = \hbar \left ( N \; \sqrt{ {k\over m} +  {\hbar^{2}
%\over 4m^{2}R^{4} }} - {\hbar \over 2mR^{2}} (N^{2}+{1\over 4})
%\right ),\;
% N= 2n'+ j +2 \; ,
%\nonumber
%\\
%\epsilon = \hbar \left ( N \; \sqrt{ {k\over m} +  {\hbar^{2}
%\over 4m^{2}R^{4} }} -  {\hbar \over 2mR^{2}} (N^{2}+{1\over 4})
%\right ),\;
% N=  2n'' + j +1\; .
%\label{73}
%\end{eqnarray}

\section{Summary}

Thus, we have discussed the behavior of a quantum-mechanical particle with spin 1 in the field of a magnetic charge using the relativistic Duffin-Kemmer-Petiau equation. We have separated the variables using the technique of the $D$-Wigner functions. In this description, there appear three quantum numbers standing for the energy, the square and the third projection
 of the generalized total angular momentum.
  The treatment results in a complicated system of 10 radial equations, the analytic solution of which is not known. The nonrelativistic approximation further reduces the problem to a system of three second-order differential equations for three radial functions.

For the Minkowski space, the equations of this system are disconnected by means
 of a linear transformation which diagonalizes the mixing matrix, and the
 problem is reduced to three ordinary differential equations of the same
 structure. Each of these equations contains as a parameter a root of a
  cubic equation appearing in bringing the mixing matrix to the
  diagonal form. The analysis is generalized to include spherically symmetric
  external fields. The cases of the Coulomb field and the external oscillator potential
  are studied. The exact solutions in terms of the hypergeometric functions
  are constructed, and for each case, three series of energy levels are derived.

For the spin 1 particle in the Lobachevsky geometry background, the three equations
of the nonrelativistic approximation are not disconnected in the presence of a monopole.
However, for the minimum
total angular  momentum, in some cases the nonrelativistic equations are solved,
for instance, for the Coulomb or oscillator potentials. For the latter two cases,
 we have constructed the solutions, in terms of the hypergeometric functions, and
  have derived the corresponding energy spectra.

Finally, we have considered the spin 1 particle in the absence of the monopole
field background. We have shown that for the Coulomb or oscillator potentials
the problem is reduced to a system of three equations involving a hypergeometric
and two general Heun equations. Though the current theory of the Heun
 equation (which presents a direct generalization of the hypergeometric equation)
 does not suggest simple algorithms for derivation of the corresponding spectra,
  we have determined the energy levels of the system by imposing on the parameters of
   the Heun equation a special requirement, which seems to be rather reasonable from the
   physical point of view.

All results obtained for the Lobachevsky  model can be easily extended to
the its geometrical counterpart, the spherical Riemann space. Due to compactness of the spherical space
all energy spectra in this case will be discrete.

\section{Acknowledgment}

This  work was   supported   by the Fund for Basic Researches of Belarus,
 F 13K-079, within the cooperation framework between Belarus  and Ukraine,
 and by the Fund for Basic Researches of Belarus,
 F 14ARM-021, within the cooperation framework between Republic of Belarus  and  Republic of Armenia.
This research has been supported by the Armenian State Committee of Science (SCS Grant No. 13RB-052) and  has been partially conducted within the scope of the International Associated Laboratory IRMAS (CNRS-France \& SCS-Armenia).

\end{document}